\documentclass[sigconf]{acmart}
\usepackage{enumitem}
\usepackage{multirow}
\usepackage{subfigure}
\AtBeginDocument{%
  \providecommand\BibTeX{{%
    \normalfont B\kern-0.5em{\scshape i\kern-0.25em b}\kern-0.8em\TeX}}}

\begin{document}

\title{A Robust Hierarchical Graph Convolutional Network Model for Collaborative Filtering}


\author{Shaowen Peng}
\email{swpeng95@gmail.com}
\affiliation{%
  \institution{Dept. of Advanced Information Technology, Kyushu University, Japan}
}
\author{Tsunenori Mine}
\email{mine@ait.kyushu-u.ac.jp}
\affiliation{%
  \institution{Dept. of Advanced Information Technology, Kyushu University, Japan}
}

\begin{abstract}
Graph Convolutional Network (GCN) has achieved great success and has been applied in various fields including recommender systems. However, GCN still suffers from many issues such as training difficulties, over-smoothing, vulnerable to adversarial attacks, etc. Distinct from current GCN-based methods which simply employ GCN for recommendation, in this paper we are committed to build a robust GCN model for collaborative filtering. Firstly, we argue that recursively incorporating messages from different order neighborhood mixes distinct node messages indistinguishably, which increases the training difficulty; instead we choose to separately aggregate different order neighbor messages with a simple GCN model which has been shown effective; then we accumulate them together in a hierarchical way without introducing additional model parameters. Secondly, we propose a solution to alleviate over-smoothing by randomly dropping out neighbor messages at each layer, which also well prevents over-fitting and enhances the robustness. Extensive experiments on three real-world datasets demonstrate the effectiveness and robustness of our model.
\end{abstract}

\ccsdesc[500]{Information systems~Recommender systems}

\keywords{Recommender Systems, Collaborative Filtering, Graph Convolutional Networks}


\maketitle

\section{Introduction}
Personalized recommender systems, which analyse user interest based on historical records to provide users with the items they might be interested in, have been playing important roles in people's daily life. There are various recommendation techniques according to different ways of inferring user preferences, most of which can be considered as extensions of matrix factorization (MF) \cite{koren2009matrix}. The MF model considers users' historical records as a user-item interaction matrix and characterizes users and items as latent vectors in a latent space; the unobserved rating is estimated as the inner product between a user latent vector and an item latent vector. Inspired by the MF model, deep learning based methods showed up and have received much attention in recent years, including the field of collaborative filtering \cite{he2017neural,xue2017deep}, sequential recommendation \cite{chen2018sequential,tang2018personalized}, recommendation with side information \cite{cheng20183ncf,feng2017poi2vec} etc. In a word, deep learning based methods assume the user-item relation is so complex that can not be sufficient to be modelled by the MF model which is considered as a linear model and employ deep learning techniques to capture non-linear user-item relations.\par

Despite the great success of above methods, the ignorance of neighbor information that can facilitate the representation of user preference limits the performance. This information includes both items that the user may be interested in and users that share the similar interests, which is similar to neighborhood-based methods \cite{sarwar2001item}. However, most neighborhood-based methods only consider the items that users have interacted with as neighbors, while they ignore the items that users have not interacted with may also contribute to user/item representations; what's more, the lack of model parameters of neighborhood-based models obstructs the models from embedding useful user/item characteristics to model user preference. To overcome the drawbacks of above methods, much attention has been attached to graph embedding techniques. Graph embedding methods combine graph models and factorization models by considering user-item interactions as a bipartite graph and encoding each node as a high dimensional vector in a vector space. We can define the directly connected nodes (the user has interacted with) as the first order (1-hop) neighbors of the target node, similarly the $n$-th order (n-hop) neighbors are defined as the 1-hop neighbors of its $(n\mbox{-}1)$-hop neighbors. Note that $\{2\cdots n\}$ neighbor connections are not included in user-item interactions even though they are close to the target node in the graph. The main difference between graph embedding methods and traditional methods is by considering user-item interactions as a graph, graph embedding methods are able to incorporate higher-order neighbor connections which are not included in original user-item interactions.\par
Amongst various graph embedding techniques, graph convolutional network (GCN) is one of the most popular ones. Unlike convolutional neural networks, GCN deals with non-Euclidean data such as graph-structured data by defining the graph convolution \cite{bruna2014spectral}. \cite{defferrard2016convolutional} reduces the computational complexity greatly by introducing the Chebyshev polynomial to simplify the graph convolution operation; \cite{kipf2017semi} reformulates \cite{defferrard2016convolutional} with $K=1$ (first-order proximity) and adds renormalization trick to alleviate gradient vanishing/exploding, the higher-order proximity is fulfilled by stacking multiple layers. The architecture of \cite{kipf2017semi} is similar to the deep neural network, which recursively repeats the propagation rule: node aggregation and non-linear transformation; the difference from the deep neural network lies in the node aggregation step, which aggregates the higher-order neighbor structure into the network to learn the graph structure. Due to the effectiveness and usefulness, GCN has been widely applied in different fields such as computer vision \cite{qi20173d}, text classification \cite{yao2019graph}, chemistry \cite{duvenaud2015convolutional} etc.\par 

However, we argue that unreasonable designs of GCN increase the training difficulty and limit the performance. Firstly, GCN recursively aggregates messages from different order neighborhood and mixes messages from distinct nodes indistinguishably, which makes it hard to extract important messages and to wash out the irrelevant messages \cite{pei2020geom}. Secondly, neighbor messages from different order neighborhood are all compressed to a fixed-length vector for the node representations, we argue that a fixed-length vector is not sufficient to faithfully represent information of the graph structure, which may cause a loss of important information in transmission. Thirdly, stacking multiple layer leads to a over-smoothing effect which degrades the performances \cite{kipf2017semi,li2018deeper}; due to this problem, most current GCN architectures remain shallow. Furthermore, recent studies \cite{zugner2018adversarial,zhu2019robust} show that GCNs are vulnerable to adversarial attacks, which is a challenge need to be tackled. In this paper, we aim to build a robust GCN model by tackling above problems for collaborative filtering. Firstly, we build our model in a hierarchical way by separating the messages from different order neighborhood. Particularly, instead of mixing all node messages together, we assume that different order neighborhoods explicitly contribute to the node representation, thus we separately aggregate neighbor messages from different order neighborhood, and then represent them in a hierarchical way for the final node representations. Secondly, we theoretically show the reason of over-smoothing and propose a solution by randomly dropping out edges of the interaction graph to slow down the convergence rate of over-smoothing, and we show that this strategy also prevents over-fitting and enhances the robustness of GCNs as well. Extensive experiments on three real-world datasets demonstrate the effectiveness and robustness of our proposed model.

\section{Related work}
\subsection{Traditional Methods}
According to the ways of dealing with user-item interactions, we can roughly identify two types of methods. The first category considers <user, item, rating> tuples as training data to model user-item relations. The MF model \cite{koren2009matrix} is the most standard one in this type, which estimates ratings as the dot product between a user latent vector and an item latent vector. Similarly, most embedding based methods fall into this category. \cite{he2017neural} points out that MF is simply a linear model which is not sufficient to model complex user-item relations and employs a multilayer perceptron to increase the non-linearity. \cite{cheng20183ncf} argues that most methods overlook the fact that users focus on different aspects on different items, and to tackle this it employs attention mechanisms to dynamically model to which aspects users pay attention. \cite{covington2016deep} uses a deep learning architecture and incorporates rich users' personal information to precisely suggest videos users may be interested in. In a word, the training strategy in this type of methods are mostly the same, the difference is the way of representing users/items in a more reasonable way.\par
The second category models consider users/items as more integrated structures. Neighborhood-based methods\cite{sarwar2001item} usually describe users as a set of items they have interacted with, and the unobserved rating is estimated based on the similarities between the target item and items the target user has interacted with. \cite{kabbur2013fism} combines MF and neighborhood-based methods by parameterizing items as latent vectors, and the similarity is calculated as the dot product between two items.  Most graph-based methods \cite{yildirim2008random,gori2007itemrank} employ random walks or markov chains to spread user preference, which works well on sparse datasets. Sequential recommendations consider user behaviors as a set of items the user has interacted with in chronological order, early studies employ markov chains \cite{rendle2010factorizing} or hidden markov models \cite{sahoo2012hidden} to tackle this problem. Some recent studies exploit RNN (and RNN variants)  to model long term temporal dependencies for predicting the user's next action; while other methods \cite{kang2018self} use attention mechanism to dynamically model the contribution of an item to user preference. In conclusion, this type of methods pay attention to a big picture to predict user interest, instead of simply focusing on the <user,item> pair; however, because of the lack of model parameters, most of the early studies are inferior to the first ones.

\subsection{Graph Embedding-based methods}
Except for GCN models, a lot of graph embedding methods also show great potentials in recent years. DeepWalk \cite{perozzi2014deepwalk} is one of the earliest studies on graph embeddings. It adopts a truncated random walk to sample the paths in a graph, the node representations are fulfilled by estimating the likelihood of the paths by employing a skip-gram model \cite{mikolov2013distributed} which has been extensively used for natural language processing to learn the context information of the nodes. \cite{cao2016deep} propose a random surfing model to capture graph structural information from a PPMI matrix, then a stacked denoising autoencoder is used to learn low-dimensional vertex representation. \cite{yang2018hop} augments the training data by employing a random walk to sample a set of items that are close to the target user in the graph, then a factorization model is introduced to learn the user-item interactions, including the items that are not directly connected to the user. GAT \cite{velickovic2018graph} employs self-attention \cite{vaswani2017attention} to dynamically model the contributions of each neighbor and aggregates all first-order neighbors for node representations of target nodes.

\subsection{GCN-based methods}
GCN has received much attention in recommender systems. \cite{berg2017graph} stacks a graph convolution layer followed by a dense layer to accumulate the messages that are aggregated according to different types of edges as the node representations, but it basically only considers the first order neighbors, which ignores the contributions from higher-order neighbors.  \cite{ying2018graph} is an extension of \cite{hamilton2017inductive} which combines random walks and graph convolutions to generate node representations; unlike vanilla GCN, it employs random walks to sample neighbor nodes to generate new embeddings, and the node representations for the next layer are generated by sending the concatenation of new embeddings and current representations to a neural network, which is computationally costly. \cite{wang2019neural} employs vanilla GCN for collaborative filtering by stacking multiple layers to incorporate higher-order connectivity, the node representation is the concatenation of the messages from different layers. \cite{chen2020revisiting} exploits a linear GCN \cite{wu2019simplifying} to reduce the model complexity and proposes a residual learning strategy to tackle the over-smoothing problem. Despite the superior performance the above-mentioned methods achieve, they still suffer from some problems because of the unreasonable designs of GCN which limit the performance. Besides collaborative filtering, GCN has also extensively applied in social recommendation \cite{fan2019graph}, knowledge-based recommendation \cite{wang2018ripplenet}, etc.

\section{Methodology}
\subsection{Problem Statement}
Given the user-item interaction matrix $R$ including $M$ users and $N$ items, the corresponding bipartite graph $\mathcal{G}=(\mathcal{V},\mathcal{E})$ includes $|\mathcal{V}|=M+N$ nodes and $|\mathcal{E}|$ edges where two nodes are directly connected only if $R_{ij}\neq 0$. The adjacency matrix of $\mathcal{G}$ is denoted by $A$, where $A_{ij}$ shows the status of the connection between node $i$ and $j$; $D$ denotes the diagonal degree matrix where the diagonal elements represent node degrees $d_i$. Each user $u$ (and item $i$) is encoded as an embedding vector $e_u\in\mathbb{R}^d$ ($e_i\in\mathbb{R}^d$), and embedding vectors of all nodes comprise the embedding matrix $E\in\mathbb{R}^{(M+N)\times d}$. We propose a GCN model to decode the embedding vectors by learning an interaction function, which can be formulated as follows:
\begin{equation}
f(R^{\mbox{-}}|\mathcal{G},{\rm \Theta}):\mathbb{R}^d\times \mathbb{R}^d\rightarrow \mathbb{R}^+ 
\end{equation}    
Given the model parameters ${\rm \Theta}$ and the interaction graph $\mathcal{G}$ which includes observed interactions $R^{\mbox{+}}$, we estimate unobserved interactions $R^{\mbox{-}}$. In this paper, we focus on implicit feedback, which means each edge in $\mathcal{G}$ shares the same weight.

\subsection{Graph Convolutional Networks}
\subsubsection{Architectures}
The core idea of GCN is to generate node representations by recursively incorporating messages from higher-order neigborhood, the propagation rule is defined as follows:
\begin{equation}
H^{(k+1)}=\sigma\left(\hat{A}H^{(k)}W^{(k)}\right)
\end{equation}
where the normalized augmented adjacency matrix $\hat{A}=\tilde{D}^{\mbox{-}1/2}\tilde{A}\tilde{D}^{\mbox{-}1/2}$, and $\tilde{A}=A+I$, $\tilde{D}=D+I$; here $I$ is an identity matrix and is introduced to add self-connections (a node is considered connected with itself). The propagation rule is comprised of two steps: node aggregation and transformation; node aggregation is fulfilled by the left-multiplication of $\hat{A}$ which is represented as $\hat{A}H^{(k)}$ at the $k$-th layer; $W^k$ is the weight matrix which maps the aggregated messages $\hat{A}H^{(k)}$ to an output space; $\sigma(\cdot)$ is an activation function. The initial state $H^{(0)}$ is usually comprised of node feature vectors of all nodes, and by stacking multiple layers, the node information from different higher-order neighborhood is aggregated to the target node to contribute to the node representations.\par
To better clarify how GCN works, here we introduce a variant named SGC \cite{wu2019simplifying}, which discards the non-linear activation function to reduce the computational complexity and shows excellent performance. The propagation rule is simplified as follows:
\begin{equation}
\hat{Y}=\sigma\left(\hat{A}{\cdots}\hat{A}H^{(0)}W^{(1)}{\cdots}W^{(n)}\right)=\sigma\left(\hat{A}^{n}H^{(0)}W\right)
\end{equation}
where $W=W^{(1)}\cdots W^{(n)}$. The node aggregation at the $k$-th layer is simply represented as: $\hat{A}^{k}H^{(0)}$ followed by a linear transformation. Note that adjacency matrix $\hat{A}$ shows the direct connections between two nodes, which is the first order neighborhood; similarly the power of adjacency matrix shows the connections of higher-order neighborhood (e.g., $\hat{A}^2_{ui}\neq0$ means $i$ is $u$'s second order neighbor and vice versa). Thus, the node aggregation at the $k$-th layer $\hat{A}^{k}H^{(0)}$ aggregates the messages from $k$-th order neighbors. Now it's clear that how GCN works: firstly we take the node feature matrix as the initial state, then the messages from first order neighborhood are aggregated to the target node to update the node embeddings, which is fulfilled by left-multiplication of $\hat{A}$, then we maps the updated embeddings to an output space (the next layer) with a weight matrix; by repeating the propagation rule, finally we get the node representations which take higher-order neighborhood into consideration.
\subsubsection{Problems in GCN}
However, we argue that the propagation rule of GCN is not reasonable enough, the reasons are threefold. Firstly, messages from different order neighborhood are transmitted and stored in fixed-length vectors, which is incapable of carrying the information of the graph structure, thereby leading to information loss in transmission. It is worth mentioning that this problem is similar to Seq2seq \cite{sutskever2014sequence}, which shows limited performance on long sentences because the information of input words is all stored in a fixed-length vectors; attention mechanism \cite{bahdanau2014neural} tackles this problem by accumulating all hidden states of the input sequence as the context vectors, which give the inspiration to us to propose our solution. Secondly, GCN simply mixes messages from distinct nodes up, which implies that distinct nodes follow the same distribution. We argue that this design may increase the difficulty in extracting important messages. Furthermore, stacking many layers leads to an over-smoothing effect which degrades the performance.\par

\begin{theorem}
Given the augmented normalized adjacency matrix $\hat{A}$, over-smoothing is formulated as $Rank(\hat{A}^k)\leq dim(V_{\lambda_{max}})$ when $k\rightarrow\infty$, where $dim(V_{\lambda_{max}})$ is the dimension of the eigenspace of the maximum eigenvalue $\lambda_{max}$ of $\hat{A}$.
\end{theorem}

\begin{figure*}
\begin{center}
\includegraphics[width=0.86\textwidth]{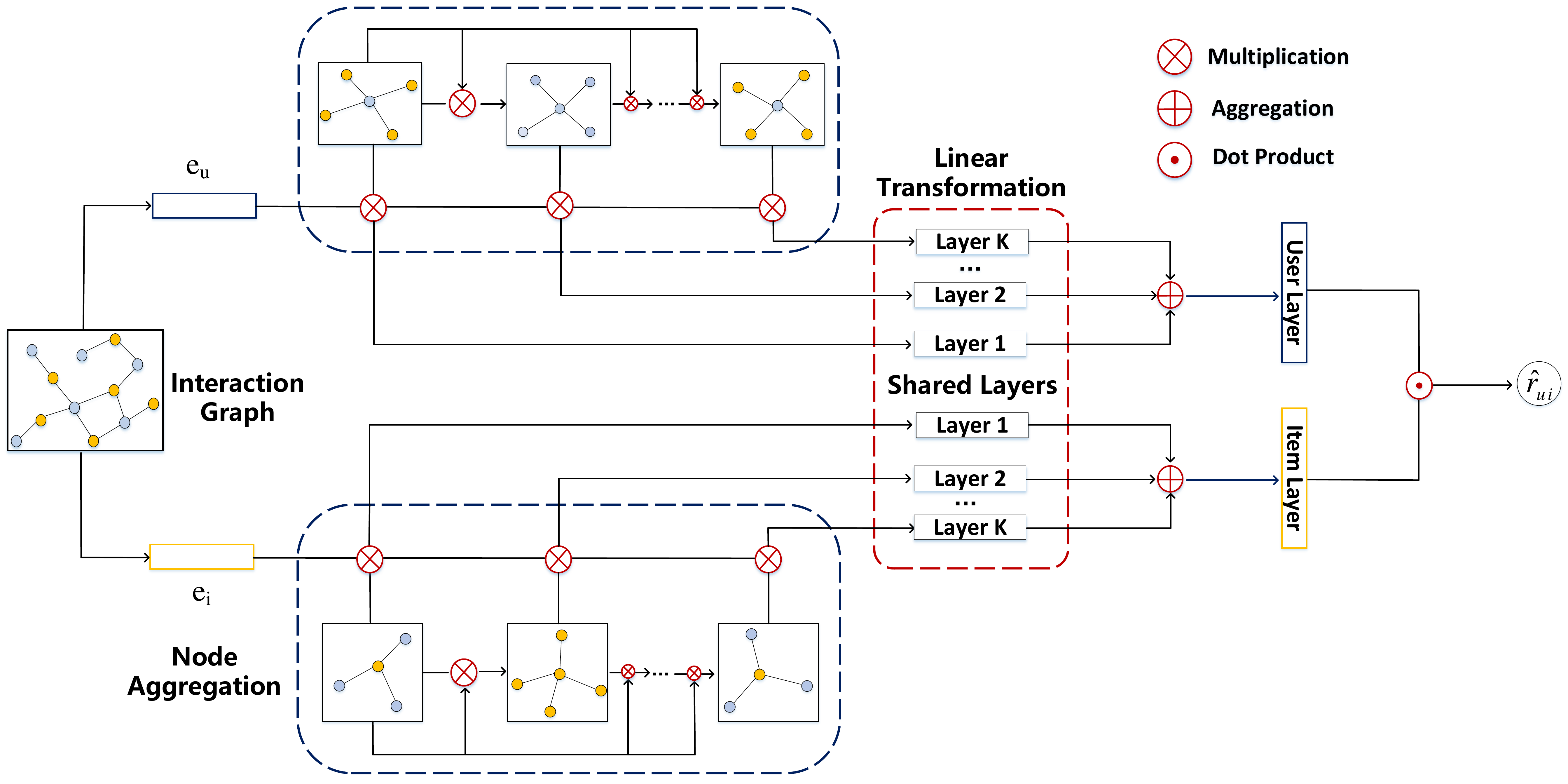}
\caption{Architecture of the proposed model.}
\label{model}
\end{center}
\end{figure*}

\begin{proof}
Following the spectral decomposition, we have:
\begin{equation}
\hat{A}^k =\sum\lambda_i^kx_ix_i^{T}
\end{equation}
where $\lambda_i$ is the $i$-th eigenvalue and $x_i$ is the corresponding orthonormal eigenvector. According to \cite{wu2019simplifying} the range of eigenvalues of the Laplacian matrix $\mathcal{L}=I-\tilde{D}^{\mbox{-}1/2}\tilde{A}\tilde{D}^{\mbox{-}1/2}$ is $\sigma_i\in[0,2)$, thus we have the following equation:
\begin{equation}
\mathcal{L}x_i=(I-\hat{A})x_i=(1-\lambda_i)x_i=\sigma_ix_i
\end{equation}      
from the relation $1-\lambda_i=\sigma_i$, it is easy to verify that the range of eigenvalues of $\hat{A}$ is $\lambda_i\in(\mbox{-}1,1]$. Note that 1 is an eigenvalue for any $\hat{A}$, where one of the corresponding eigenvectors is $\tilde{D}^{1/2}[1,\cdots,1]^T$. Therefore, when $k\rightarrow\infty$:
\begin{equation}
\hat{A}^k=\sum_i x_{max,i}x_{max,i}^{T}
\end{equation}
where $x_{max,i}$ is the eigenvector of the maximum eigenvalue $\lambda_{max}=1$. Thus, the neighborhood representation for any node can be represented as a linear combination of eigenvectors of the the maximum eigenvalue, which exactly means $Rank(\hat{A}^k)\leq dim(V_{\lambda_{max}})$.       
\end{proof} 
It is worth mentioning that, in most cases the components of $\tilde{D}^{1/2}[1,\cdots,1]^T$ account for absolute weights, thus we can consider the node neighborhood representation for a node $n$ as $(\hat{A}^k)_n\propto \tilde{D}^{1/2}[1,\cdots,1]^T$ and $Rank(\hat{A}^k)\approx 1$ when $k\rightarrow\infty$. In a conclusion, over-smoothing happens at the node aggregation step when $k>1$, and it becomes more severe as $k$ increases; eventually, neighborhood representations of distinct nodes converges to the same patterns. What's adding the fuel to the fire is that the messages from which the over-smoothing is not so severe are sent to the next layer, which mixes them up with those node messages which are already 'polluted' by the overs-smoothing, thereby leading to another problem of extracting the useful messages from the 'polluted' messages. Consequently, it is of great importance to incorporate the higher-order neighborhood in a hierarchical way to prevent different order messages from mixing together.
 
\subsection{A Solution to Over-Smoothing}
Recent studies adopts techniques for training convolutional neural networks (CNN) such as residual networks to train GCN and achieved superior performance \cite{li2019deepgcns}. However, as shown in the previous section , over-smoothing is largely due to the power of $\hat{A}$, thus we propose a strategy to slow down the convergence rate by putting random noise on $\hat{A}$. There are many ways to generate random noise, due to the concern of deformation of the graph structure if we add some noise on graph, we decide to deduct information from the graph, which has been shown effective in node classification \cite{rong2020dropedge}. Our solution is formulated as follows: 
\begin{equation}
A_{drop}=\hat{A}\otimes B 
\end{equation}
\begin{figure}
\begin{center}
\includegraphics[width=0.3\textwidth]{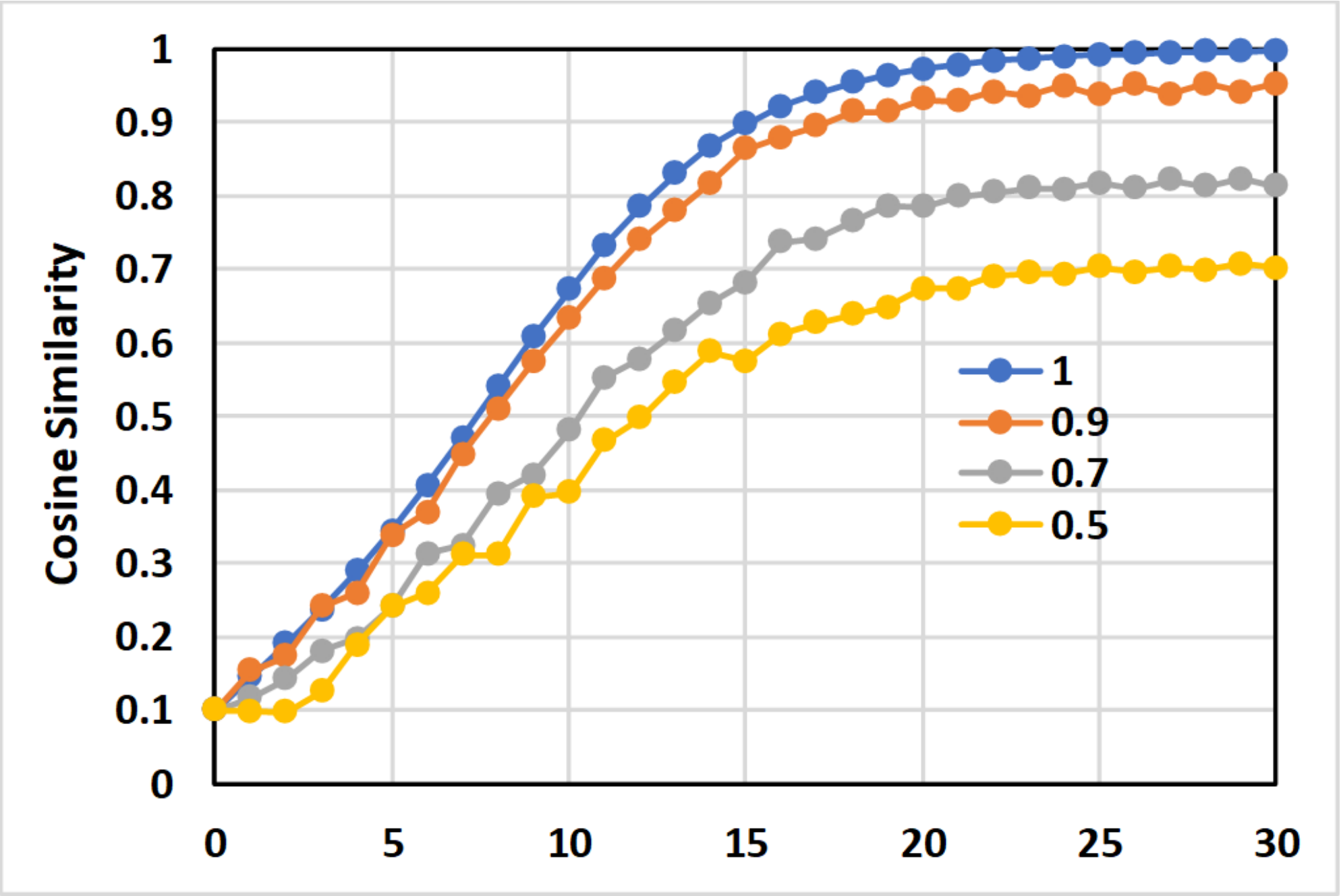}
\caption{Convergence rate of $\hat{A}$ with varying values $p$. X-axis is the $k$ in $A_{drop}^k$, y-axis shows the cosine similarities between current node vectors of $A_{drop}^k$ and the node vectors when they completely converges, which can be considered as the convergence ratio $\in [0,1]$.}
\label{over-smoothing}
\end{center}
\end{figure}
where $B$ is a binary matrix which has the same size with $\hat{A}$, and the element $b_{ui}\sim Bernoulli(p)$; $\otimes$ denotes an element-wise multiplication. For $\hat{A}_{ui}$, $b_{ui}$ is an element of independent Bernoulli random variables each of which has probability $p$ of being 1; thus there is a probability of $1\mbox{-}p$ that $\hat{A}_{ui}$ is dropped out, and the dropped out messages are not aggregated to the target node. $B$ can also be designed as a binary vector for saving the memory, since $A$ is usually a large sparse matrix, which is represented as follows:
\begin{equation}
A=\begin{bmatrix}
 0& R\\ 
R^T &0 
\end{bmatrix}
\end{equation}
where $0$ is a zero matrix. We can see that most of the elements are $0$ which do not need to be dropped out. The equation (7) is performed for each multiplication of $\hat{A}$, and we simply denote by $A_{drop}$; note that $A_{drop}$ is different at different layers.\par
Theoretically speaking, by putting sufficiently intense perturbations on the adjacency matrix at different layers, equations (4) and (5) do not holds any more since $A_{drop}^k\neq \hat{A}^k$, thus the over-smoothing no longer exists. To verify this assumption, we conduct experiments on a public dataset FilmTrust \cite{guo2013novel} with 35497 interactions. The results are shown in Figure \ref{over-smoothing}. We can see that the convergence rate at $p=1$ is extremely quick, where it completely converges at $k=20$; while the curve at $p=0.9$ shows a slower convergence rate and almost stops at 0.9; the same trend shows at $p=0.7$ and $p=0.5$ with smaller convergence ratios. In a nutshell, by randomly dropping out graph edges, the over-smoothing problem can get well alleviated. However, there is a concern that a large drop ratio causes the graph structure to be consistently distorted, which does not necessarily lead to a better performance. Therefore, there is a trade-off between the containment of the over-smoothing problem and the integrity of graph structure; small drop ratios can not well contain over-smoothing, while large drop ratios lead to a incomplete graph structure; thus an appropriate value is needed. We will discuss the settings of drop ratios in details in the later section.\par
Furthermore, we argue that our solution can effectively prevent over-fitting. GCN is basically designed for transductive learning and put the information of the whole graph structure into training; while since we focus on inductive learning and the interaction graph $\mathcal{G}$ only includes training interactions, thus consistently feeding the same incomplete graph into training time leads to over-fitting. Distinct from GCN, $A_{drop}$ can be considered as a sub-graph of $\mathcal{G}$, i.e., we sample different sub-graphs for training every time, which helps the model better comprehend the graph structure and is beneficial to the generalization. On the other hand, randomly dropping out edges can be considered as imperceptible noise consistently imposed on the graph, which is expected to enhance the robustness of GCN.
\subsection{Proposed Model}
To tackle the problems mentioned in section 3.2 we propose our model named RH-GCCF, which is illustrated in Figure \ref{model}. We elaborate the architecture step by step.\\
\textbf{Propagation.} Given the initial state $H^{(0)}$, node embeddings incorporating different order neighborhood are generated as follows:
\begin{equation}
H^k=A_{drop}^kH^{(0)}W_k
\end{equation}
Here $H^{(0)}=E$, and we replace $\hat{A}$ with $A_{drop}$; $W_k$ is a linear transformation where $W_k=W^{(1)}\cdots W^{(k)}$; note that we do not use non-linear activation function here, so multiplications of multiple weight matrices can be seen as a linear transformation. Equation (9) considers updates with the matrix form, which is equivalent to the following equation when we consider each user $u$ and item $i$:
\begin{equation}
h^k_u=\left[a^k_{uu}e_u+\sum_{j\cap a^k_{uj}\neq 0}a^k_{uj}e_j \right]W_k
\end{equation}
\begin{equation}
h^k_i=\left[a^k_{ii}e_i+\sum_{m\cap a^k_{im}\neq 0}a^k_{im}e_m \right]W_k
\end{equation}
where $a^k_{uj}$ is the coefficient of $A_{drop}^k$; $a^k_{uj}\neq 0$ shows the edge connections that have not been dropped out. Note that when $k=1$, $a_{uu}=1/d_u$, $a_{uj}=1/\sqrt{d_u\times d_j}$.\par   
We can see that the propagation rule in our model is similar to SGC, which is introduced in section 3.2, the difference is that SGC still adopts the neural network architecture that recursively sends information in the current layer to the next layer; while we choose to directly output the node embeddings of different order neighborhood to the final layer, which prevents messages from different order neighborhood from mixing up and also prevents the clean messages from being 'polluted' by the message that are deeply affected by over-smoothing.\\\\
\textbf{Prediction.} After we get the node embeddings from different order neighborhood, we user an aggregation function $aggre(\cdot)$ to aggregate them together:
\begin{equation}
O=aggre\left(H^{(0)},H^{(1)},...,H^{(K)} \right)
\end{equation}
There are several choices such as max pooling, LSTM, etc., which have been applied in recent work \cite{hamilton2017inductive}. To avoid introducing additional model parameters, there are two choices: sum function and concatenate function. In our model, we choose to use concatenate function to generate the final node representations in that it enable the model to represent the features of different order neighborhood in a hierarchical way. Some recent studies \cite{wang2019neural} adopt the similar strategy with the vanilla GCN to represent node representations, however, because of the non-linear activation function, neighbor messages from different order can not be generated independently like we do $H^k=A_{drop}^kH^{(0)}$, instead they have to take the messages from previous layer to generate node embeddings as shown in equation (2), which mixes the messages from distinct node indistinguishably and introduces unnecessary dependencies which increase the training difficulty.\par
Finally, the interaction between a user $u$ and an item $i$ is predicted as follows:
\begin{equation}
\hat{r}_{ui}=o_u^To_i
\end{equation}
where $o_u$ and $o_i$ are the corresponding node vectors from $O$.

\begin{figure}
\begin{center}
\includegraphics[width=0.28\textwidth]{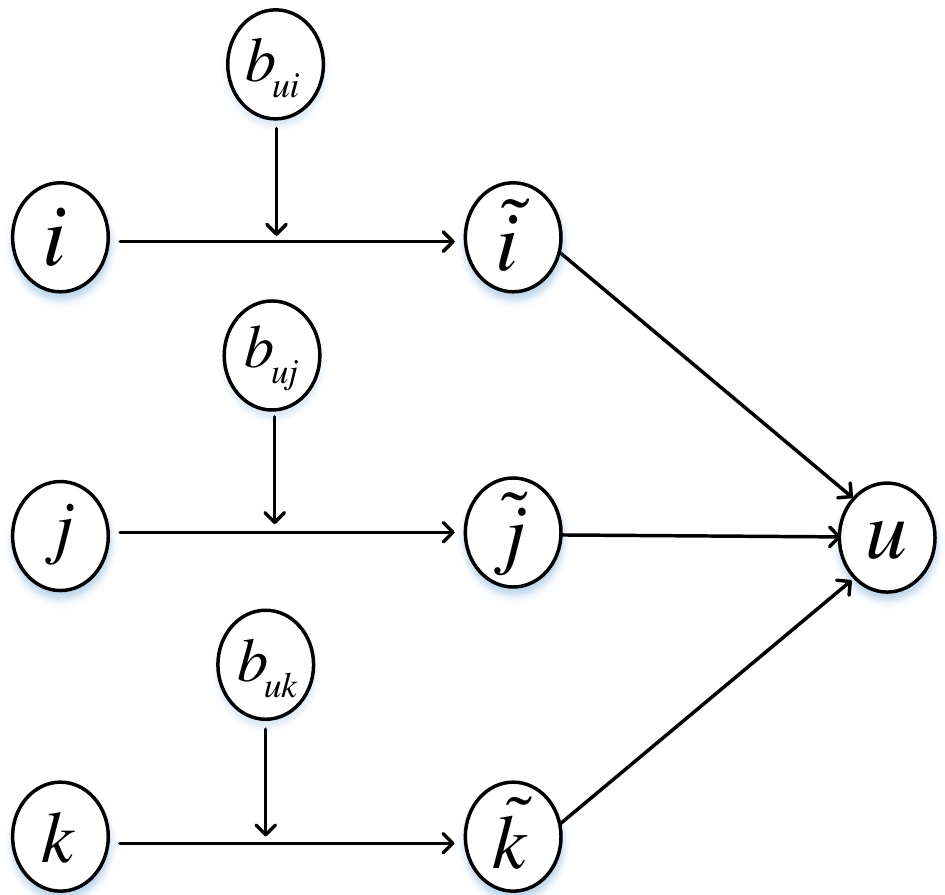}
\caption{A visualization of how node messages get dropped out}
\label{dropout}
\end{center}
\end{figure}

\subsection{Discussion}
\subsubsection{Comparison with Dropout} 
We found our idea shares a lot of similarities with Dropout \cite{srivastava2014dropout} which is designed to prevent the over-fitting of neural networks. The core idea of Dropout is to randomly drop out the neurons with a certain ratio at each layer. Even though directly employing Dropout for GCN is of no help to the containment of over-fitting or over-smoothing, but when we replace the notations in the illustration of the Dropout network with ours, we found it can perfectly explain our idea, which is shown in Figure \ref{dropout}. Instead of considering them as neurons of a neural network, we consider them as nodes in a graph; the messages from neighbors $i$, $j$, $k$ are aggregated to the target node $u$, while in the meantime binary variables are introduced to determine to let them go or drop them out. In other words, we are technically doing the same thing to prevent over-fitting, the difference is that Dropout focuses on neural networks, and we focus on graph structures.  
\subsubsection{Comparison with Attention Mechanism}   
Seq2seq is an encoder-decoder model for sequence learning, where the information of input sequence is recursively transmitted through the encoder (a RNN); the final hidden state of the encoder is sent to the decoder as a context vector including input sequence information. However, the fixed-length context vector is incapable of carrying all input information without loss, which limits its capacity. What's more, all input information is blended in disorder, which increases the training difficulty. Attention mechanism resolves this drawback by aggregating all hidden states of the input sequence as context vectors and defining attention scores to explicitly measure the contribution of each part of the input sequence to the each part of the output sequence. Inspired by how attention mechanism improves Seq2seq, we can see that GCN also suffers from the same problem, where neighborhood messages are mixed indistinguishably and are recursively transmitted through fixed-length vectors to generate the node representations. Analogously, we tackle this by accumulating the messages from different neighborhood without mixing them up, and the node representations clearly reflect the contributions of different neighborhood.

\subsubsection{Robustness to Adversarial Attacks}
The vulnerability of deep learning based methods has been a serve problem. Recent studies \cite{zugner2018adversarial,zhu2019robust} show that GCNs are also vulnerable to adversarial attacks because of the similar architectures to deep learning models. If we consider the corresponding graph of $A_{drop}$ as $\mathcal{G}_{drop}$, then $\mathcal{G}_{drop}$ can be seen as a corrupted version of $\mathcal{G}$ under adversarial attacks. Following the equation (1), our goal becomes to correctly estimate the unobserved ratings based on the incomplete interaction graph $\mathcal{G}_{drop}$ under the attack caused by equation (7). Instead of relying on the full neighborhood, dropping out edge connections with appropriate ratios and sending noisy (incomplete) data for training enable the model to automatically distinguish the real important neighbor messages and ignore the useless information, thereby enhancing the robustness under adversarial attacks.

\subsection{Optimization} 
We optimize model parameters with a pair-wise BPR loss function for the task of personalized ranking \cite{rendle2009bpr}, which is formulated as follows :
\begin{equation}
Loss=\sum_{(u,i,j)\in{T}}-\ln\sigma(\hat{r}_{ui}-\hat{r}_{uj})+\lambda\left \| \Theta \right \| ^2
\end{equation}
where $T=\{(u,i,j)|(u,i)\in{R^+}, (u,i)\in{R^{\mbox{-}}}\}$; $\Theta$ is the trainable parameters where $\Theta=\{\{W^{(1)}\cdots W^{(K)}\}, E\}$; $\lambda$ is the regularization parameter to prevent over-fitting; for $\sigma{(\cdot)}$ we use the sigmoid function. Unlike the point-wise loss function, BPR loss focuses more on the personalized rankings, which assumes that observed interactions shows higher preferences than the unobserved interactions, and note that each training pair is randomly generated from the training dataset $T$.      
\begin{table}
\small
\centering
\caption{Statistics of datasets}
\begin{tabular}{lcccc}
\toprule
Datasets&\#User&\#Item &\#Interactions &Density\%\\
\midrule
Pinterest&37,501&9,836&1,025,709&0.278\\
citeulike-a&5,551&16,981&210,537&0.223\\
Movielens&9,999&24,328&1,496,517&0.615\\
\bottomrule
\end{tabular}
\end{table} 
\section{Experiments}
We conduct experiments on three real-world datasets to evaluate our model. Particularly, we aim to answer the following research questions:
\begin{itemize}[leftmargin=10pt]
\item \textbf{RQ1}: How do hyper-parameters such as layers of the network, drop ratio affect the performance?
\item \textbf{RQ2}: Is there an optimal point with respect to performance balancing between the containment of the over-smoothing problem and the integrity of graph structure?
\item \textbf{RQ3}: Does our model outperform other state-of-the-art baselines?  
\end{itemize} 
\subsection{Experimental Setup}
\textbf{Datasets.} The descriptions of datasets are listed as follows. The statistics are summarized in Table 1.
\begin{itemize}[leftmargin=10pt]
\item \textbf{Pinterest}: This is an implicit feedback dataset for content-based image recommendation, which is collected by \cite{he2017neural}.
\item \textbf{citeulike-a}: This dataset \cite{wang2011collaborative} is collected from CiteULike which provides users with a service to save and share academic papers. The interactions are implicit feedbacks.
\item \textbf{Movielens}: This dataset contains movie ratings, which is collected by GroupLens $\footnote{https://grouplens.org/datasets/}$ for new research. Since it's an explicit-feedback datasets, we transform ratings to implicit feedbacks. We use a subset of the whole dataset for experiments.
\end{itemize}
\textbf{Evaluation Metrics.} To evaluate the performance of top-k recommendation task, we adopt two evaluation metrics which are extensively used for personalized ranking tasks: recall@k and ndcg@k. Recall measures the number of items in the recommendation lists also appeared in the user's test set; ndcg focuses more on the positions in which the items appeared in the test set by assigning a high score to the item in the top ranks. 
\\\\
\textbf{Baselines.} We compare our proposed RH-GCCF model with the following methods:
\begin{itemize}[leftmargin=10pt]
\item BPR \cite{rendle2009bpr}: This is a stable baseline which proposes a Bayesian pair-wise loss function to learn from implicit feedback.  
\item Neurec \cite{zhang2018neurec}: This is a deep learning based model which adopts an MLP to learn the non-linear relations between users and items. We set the neuron size at each layer to 150 and use three-layer architecture as the baseline.
\item GCMC \cite{berg2017graph}: This is a GCN-based model which originally focuses on the explicit feedback and uses different weight matrix to decode different types of edges. Since we focus on the implicit feedback which treats each edge as the same, it can be simply considered as an one-layer GCN model.    
\item NGCF \cite{wang2019neural}: This is a state-of-the-art GCN-based model which can be considered as an extension of \cite{kipf2017semi}. According to reports on the performance in the paper, we use three-layer architecture as the baseline.
\item LR-GCCF \cite{chen2020revisiting}: This model exploits \cite{wu2019simplifying} and proposes a residual network structure to tackle the over-smoothing problem. According to reports on the performance in the paper, we use three-layer architecture as the baseline. 
\end{itemize}
\begin{figure}
\begin{center}
\includegraphics[width=0.26\textwidth]{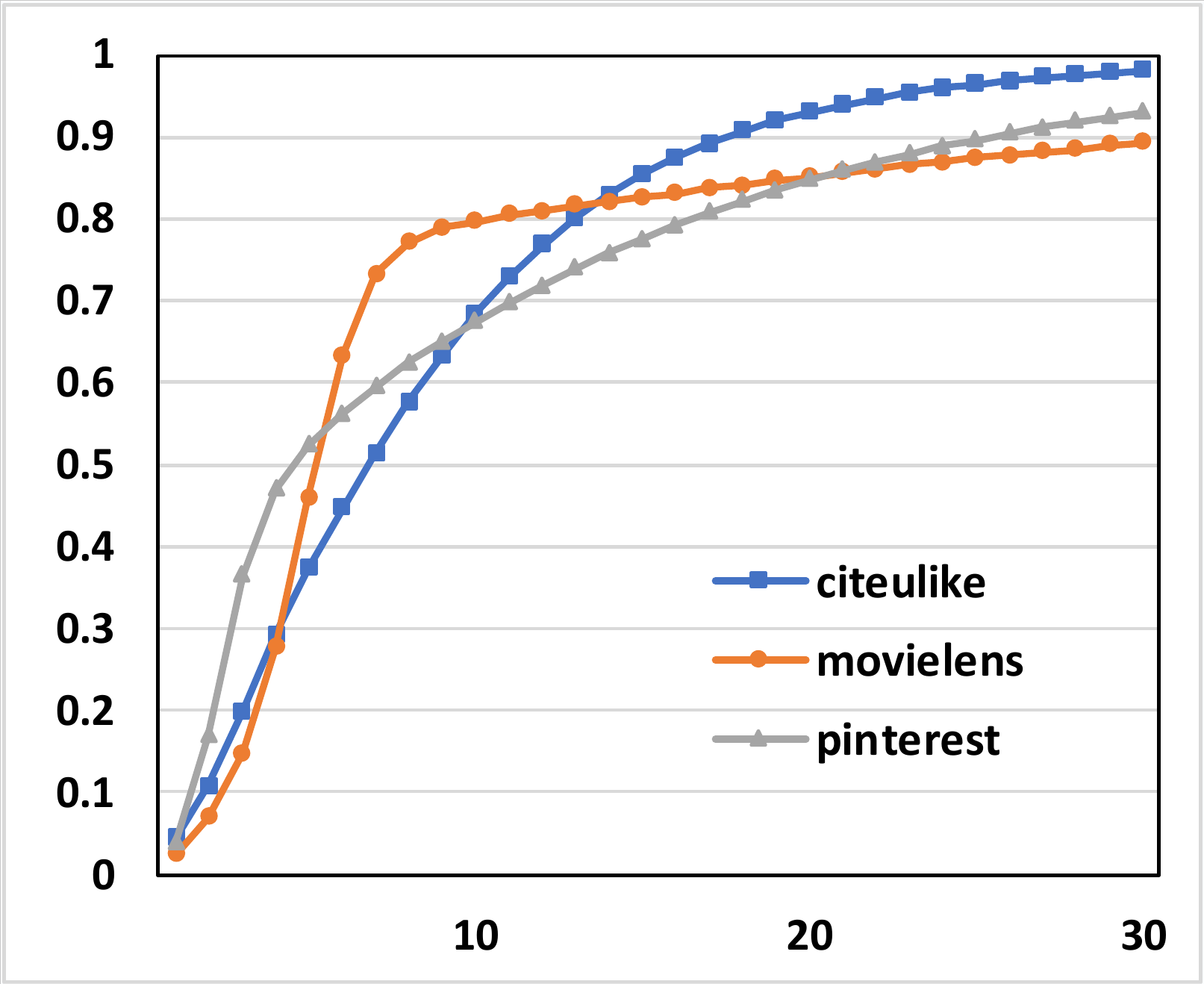}
\caption{Visualization of how convergence rate changes on three datasets. X-axis is the k in $\hat{A}^k$, y-axis represents the convergence ratio.}
\label{convergence}
\end{center}
\end{figure}
\textbf{Implementation details.} We implemented the proposed model based on PyTorch$\footnote{https://pytorch.org/}$, the code will be released upon acceptance. For all models, the optimizer is SGD; for all models, the embedding size is set to 64; we set the regularization rate to $10^{\mbox{-}3}$; the learning rate is tuned amongst $\{0.01,0.02,0.04,\cdots,0.2\}$; for model parameters, we initialize with Xavier Initialization. For other hyper-parameters we report the settings in the next section. We use 80\% of the user-item pairs for training data and leave 20\% for test.

\begin{figure} \centering 
\subfigure[Performance on citeulike with varying $K$ at $p=0.9$.  ] {  
\includegraphics[width=0.38\columnwidth]{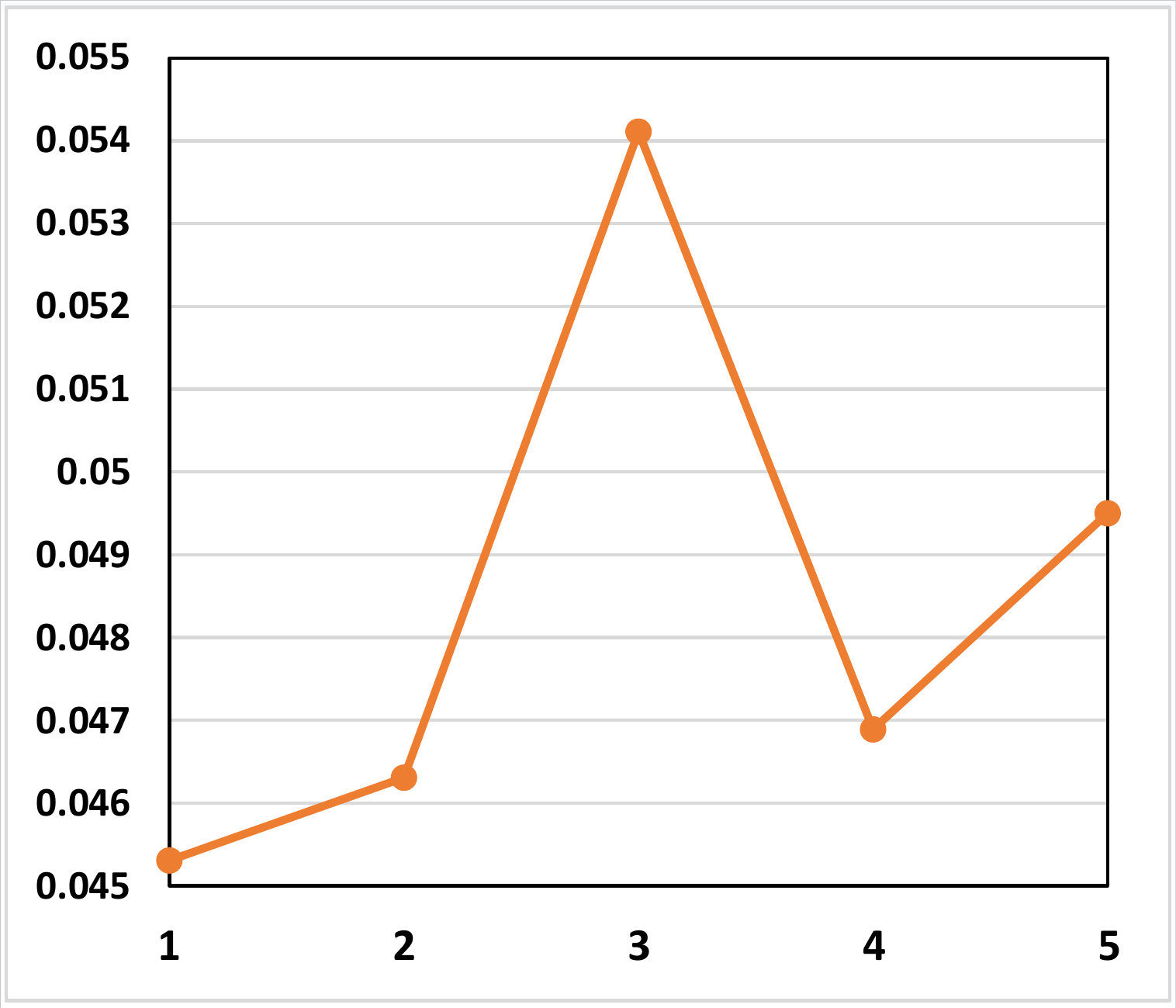} 
}  
\hspace{0.2cm}
\subfigure[Performance on Movielens with varying $K$ at $p=0.8$. ] {  
\includegraphics[width=0.38\columnwidth]{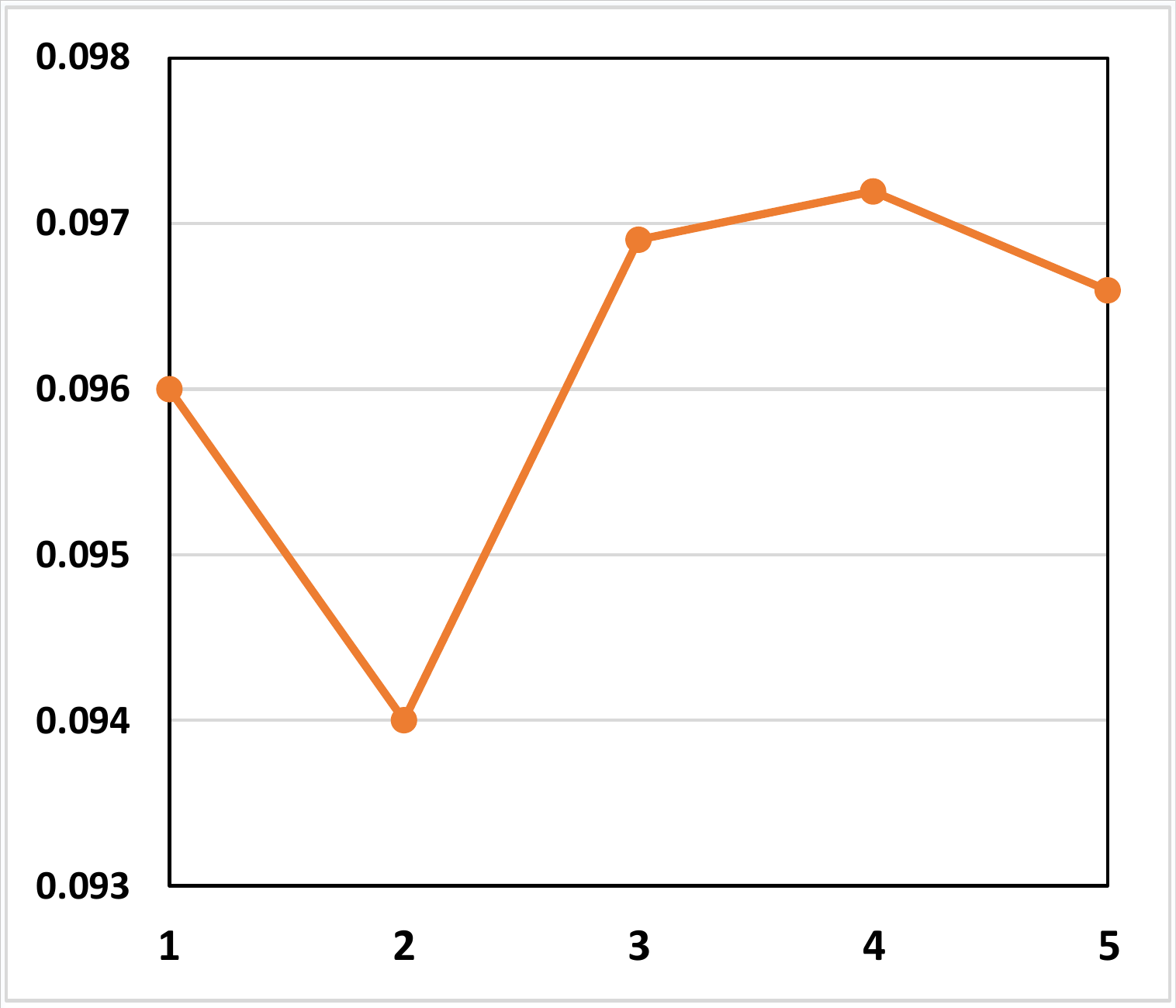} 
}
\hspace{0.2cm}
\subfigure[Performance on citeulike with varying $p$ at $K=3$. ] {  
\includegraphics[width=0.38\columnwidth]{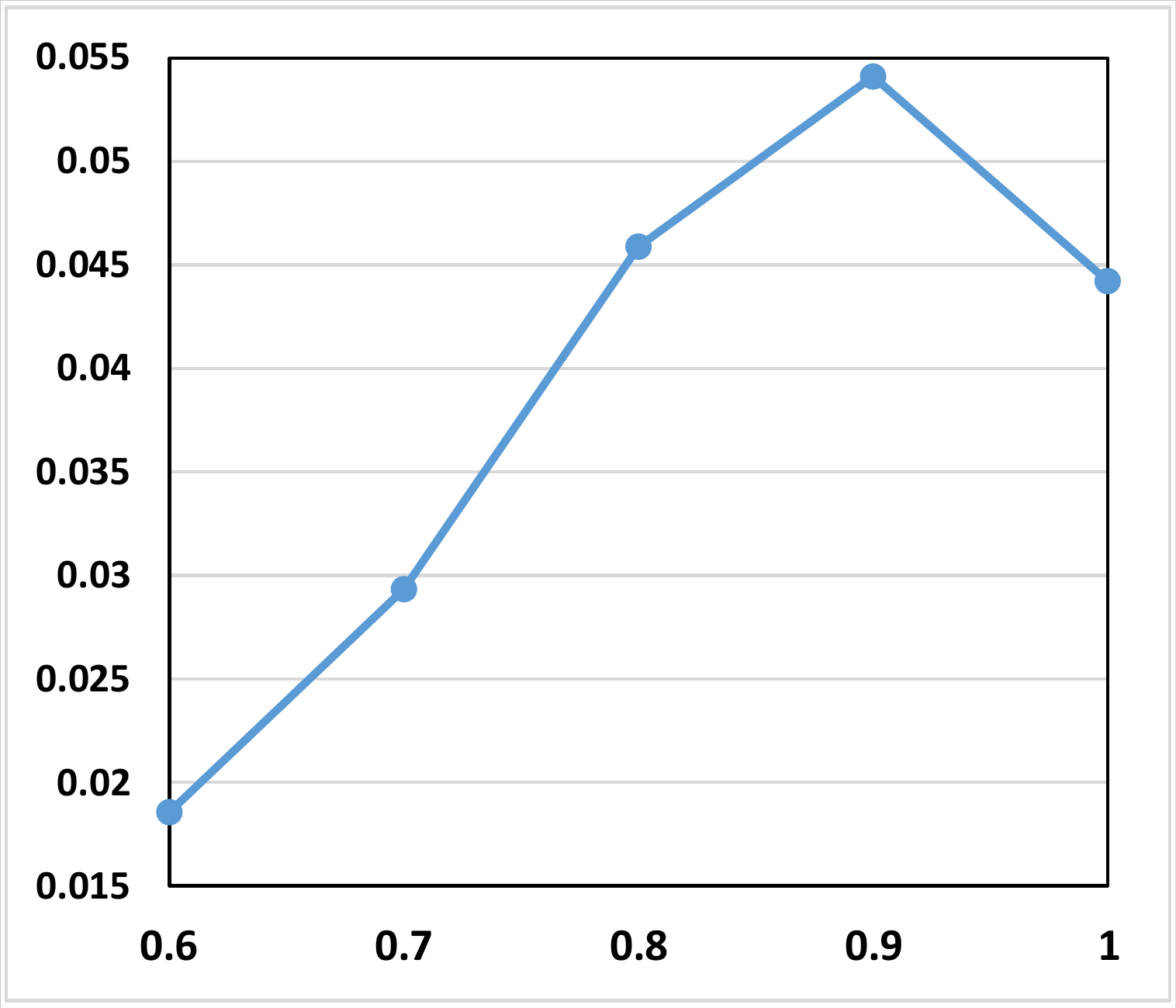} 
}
\hspace{0.2cm}
\subfigure[Performance on movielens with varying $p$ at $K=3$. ] {  
\includegraphics[width=0.38\columnwidth]{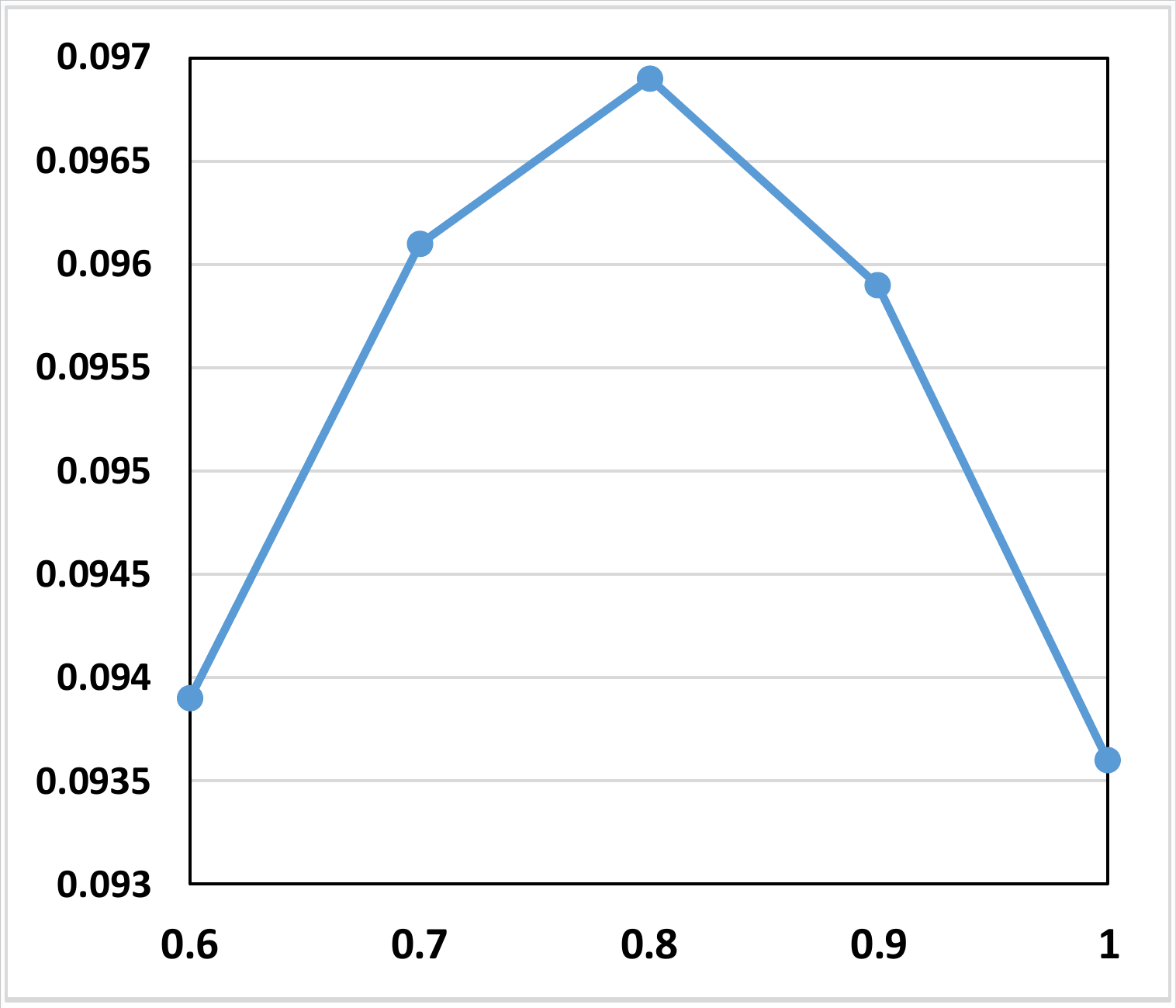} 
}   
\caption{ Effect of the drop ratio in terms of ndcg@10. }  
\label{drop_ratio}
\end{figure}

\begin{figure*} \centering 
\subfigure[Performance on Citeulike with different layers.  ] {  
\includegraphics[width=0.48\columnwidth]{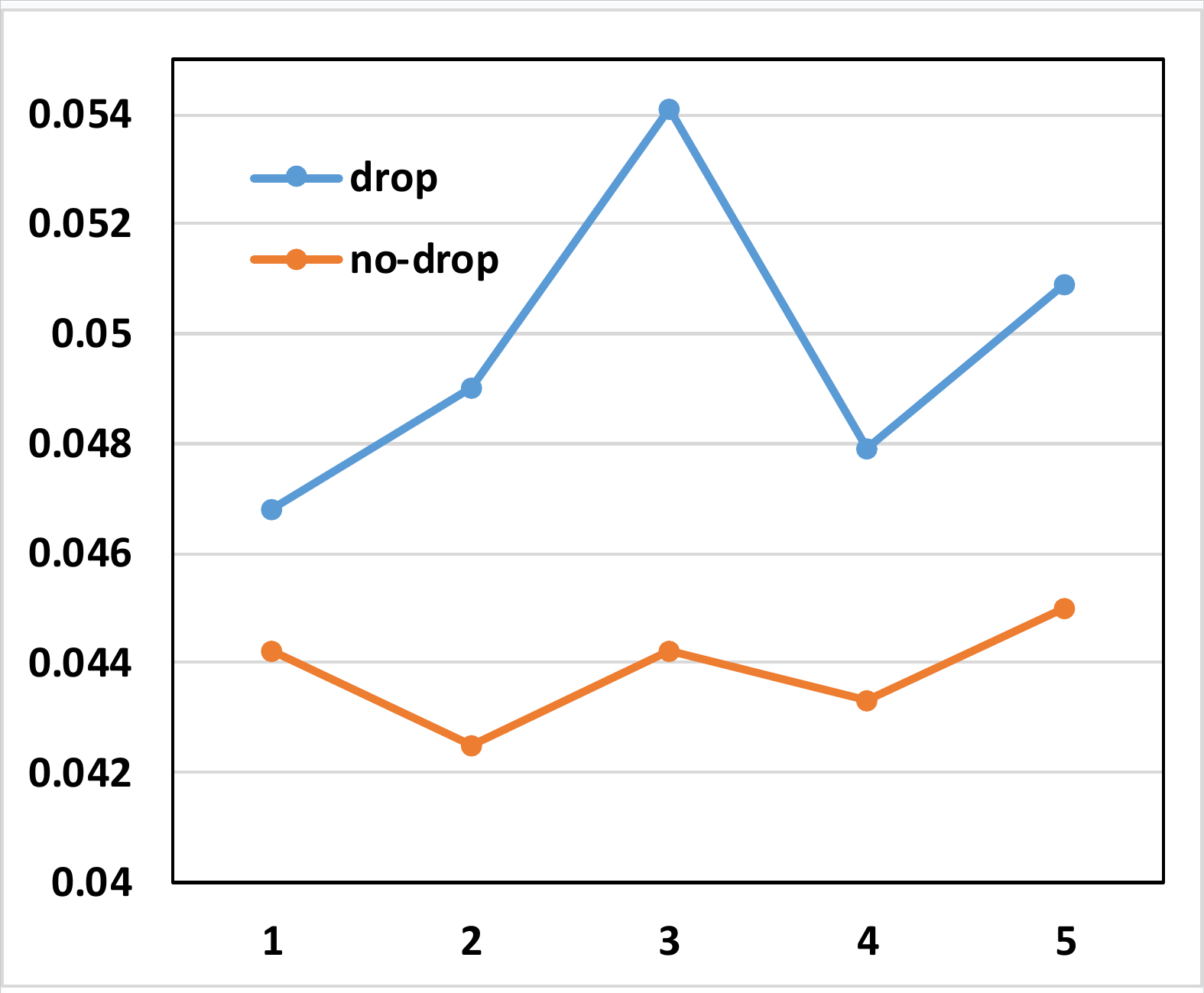} 
}  
\hspace{0.35cm}
\subfigure[Performance on Pinterest with different layers. ] {  
\includegraphics[width=0.48\columnwidth]{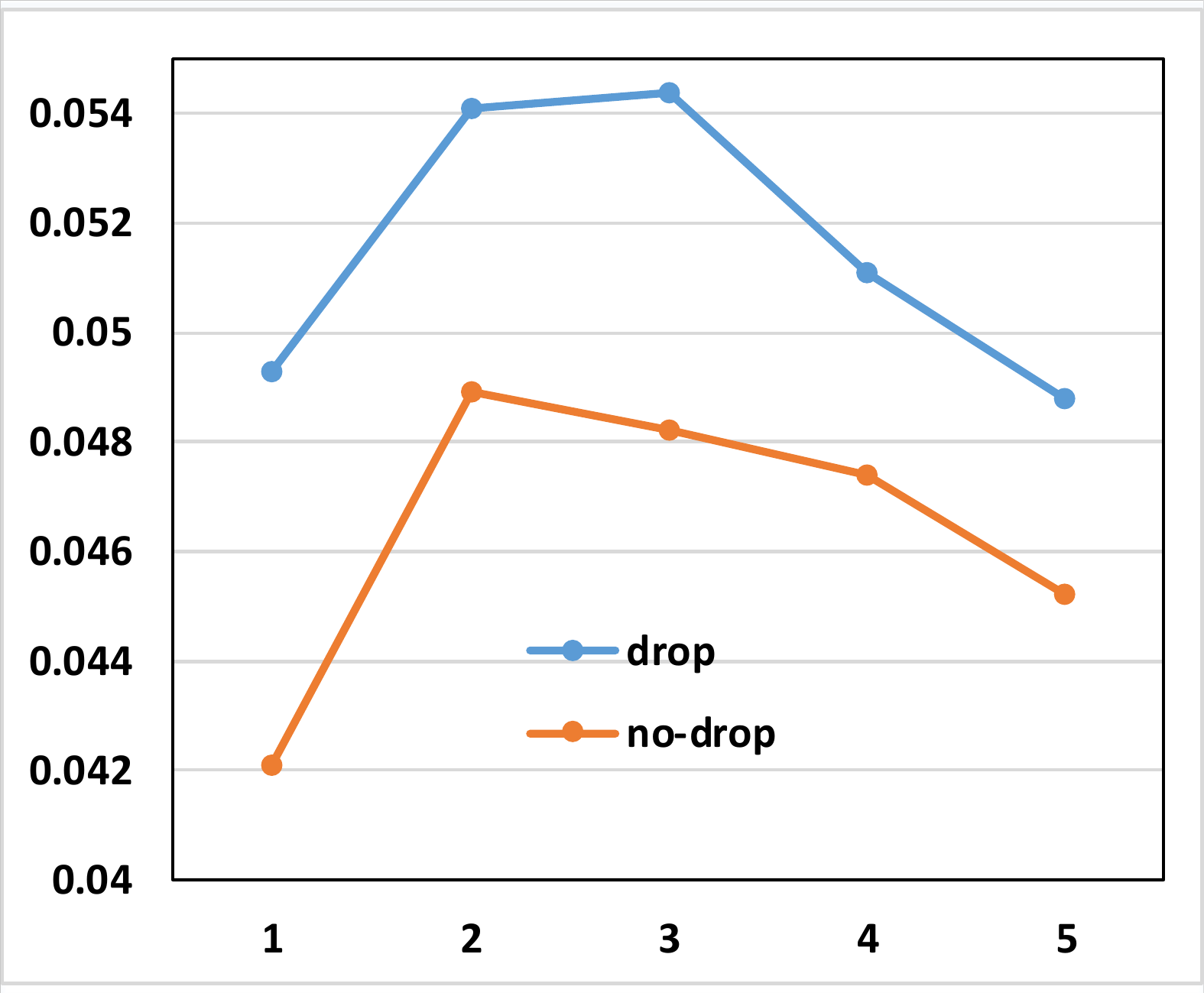} 
}
\hspace{0.35cm}
\subfigure[Performance on Movielens with different layers. ] {  
\includegraphics[width=0.48\columnwidth]{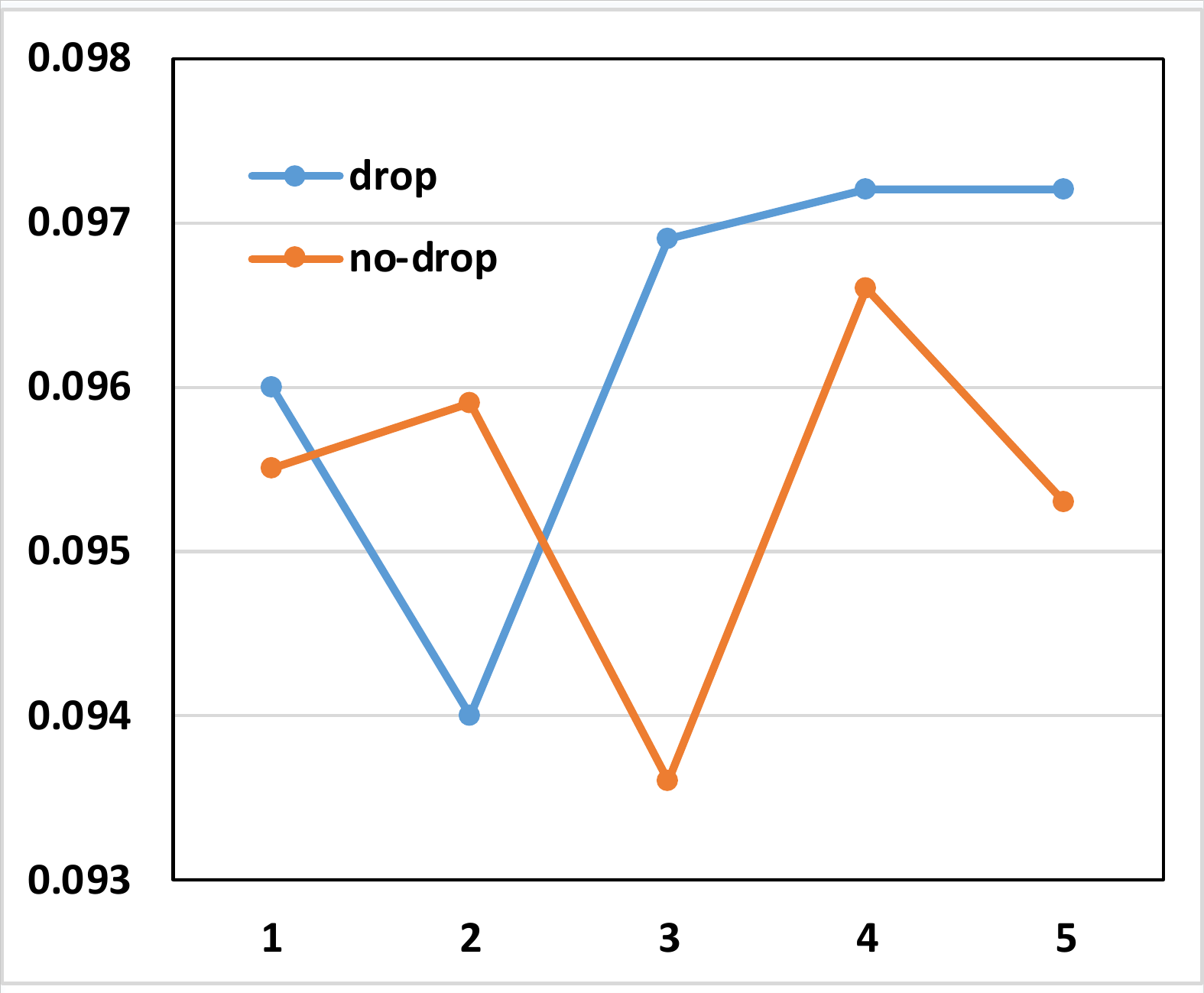} 
} 
\caption{ Effect of layer-wise noise in terms of ndcg@10. 'Drop' represents our model; 'no-drop' is the model without layer-wise noise. X-axis is the number of layers. }  
\label{layer_comparison}
\end{figure*} 
\subsection{Model Analysis (RQ1 and RQ2)} 
\subsubsection{Drop Ratios.}
We mentioned in the previous section that an appropriate value of the drop ratio is required to balance between the containment of over-smoothing and over-fitting and the integrity of graph structure. After conducting extensive experiments, we found the optimal drop ratio value indeed exists and is related to two factors: convergence rate and the layers of networks. We found that the optimal value $p$ when $K=3$ on Movielens, Pinterest, citeulike is 0.8, 0.9, 0.9, which is partially illustrated in Figure \ref{drop_ratio} (c), (d). According to the convergence rate shown in Figure \ref{convergence}, we can see that the convergence rate on Movielens is way faster than the other two; note that here we focus on $k<10$, since in most cases there is no need to build deep networks for recommender systems. We speculate that for the datasets which have a more severe over-smoothing problem (convergence rate is fast), the optimal drop ratio tends to be large (note that drop ratio is $1{\mbox{-}}p$). In the meantime, from Figure \ref{drop_ratio} (a), (b) we observe that the optimal value on the current network does not necessarily lead to a superior performance when we increase the layers, which may due to the reason that the layer-wise noise is transmitted and accumulated on the network, i.e., the $k+1$ layer takes the layer-wise noise from $\{1,\cdots,k\}$ layers, which means that messages from higher-order neighborhood suffer more from the noise than messages from lower-order neighborhood, and there is a point after which the massages are full of so much noise that can not facilitate performance. Take Citeulike as an example, we can see that the performance increases at first, where the layer-wise noise well contains the over-smoothing and over-fitting; the model maximizes the performance at $K=3$ and starts to drop, where the accumulated layer-wise noise starts to hamper the messages to be precisely transmitted and reduces the performance. The similar trend is also shown on Movielens, where the model achieves the best performance at $K=4$. 

\begin{table}[]
\caption{Performance with different layers $K$ in terms of recall@10 and ndcg@10.}
\begin{tabular}{c|c|c|c|c}
\hline
                     &        & Citeulike       & Pinterest       & Movielens       \\ \hline
\multirow{3}{*}{K=1} & recall & 0.0579          & 0.0592          & 0.1003          \\
                     & ndcg   & 0.0468          & 0.0493          & 0.096           \\
                     & $p$      & 0.9             & 0.9             & 0.8             \\ \hline
\multirow{3}{*}{K=2} & recall & 0.0594          & 0.0637          & 0.0975          \\
                     & ndcg   & 0.049          & 0.0541          & 0.094           \\
                     & $p$      & 0.9             & 0.9             & 0.8             \\ \hline
\multirow{3}{*}{K=3} & recall & \textbf{0.0648} & \textbf{0.0646} & 0.1007          \\
                     & ndcg   & \textbf{0.0541} & \textbf{0.0544} & 0.0969          \\
                     & $p$      & \textbf{0.9}    & \textbf{0.9}    & 0.8             \\ \hline
\multirow{3}{*}{K=4} & recall & 0.057           & 0.0614          & \textbf{0.1013} \\
                     & ndcg   & 0.0479          & 0.0511          & \textbf{0.0972} \\
                     & $p$      & 0.95            & 0.9             & \textbf{0.8}    \\ \hline
\multirow{3}{*}{K=5} & recall & 0.0607          & 0.0583          & 0.1011          \\
                     & ndcg   & 0.0509          & 0.0488          & 0.0972          \\
                     & $p$      & 0.95            & 0.95            & 0.9             \\ \hline
\end{tabular}
\end{table}
   
\subsubsection{Layers of the network.}
The layers of the network correspond to the order of neighboorhood. To investigate how the number of layers affects the performance, we set $K=\{1,\cdots,5\}$. The experimental results are shown in Table 2. It is obvious that higher-order neighborhood substantially increases the model performance. For example, the model with the best performance outperforms the model with only first-order neighborhood by 15.1\%, 9.1\%, 1\% on three datasets, respectively. On the other hand, constantly increasing layers of the network does not lead to a consistent improvement, which is due to the quick expansion of higher-order neighborhood. For instance, the density of $A^3$ (third-order neighboorhood) is 91.8\% on citeulike, which means that almost any two nodes are connected within three hops. Therefore, we can see that the model maximizes the performance at $K=3$ on citeulike, because keeping increasing layers only introduces the message that has been included, which causes a severe over-fitting. This is also a main reason why current GCN-based methods remain shallow. Furthermore, we observe that the optimal drop ratios tend to be smaller on deeper networks, which is consistent with the above analysis of drop ratios.
\begin{table*}[]
\caption{Overall comparison in terms of recall@k and ndcg@k where k=\{10, 20\}. The best performance is highlighted in bold, and the best baseline is underlined. The last row is the improvement of RH-GCCF over the best baseline.}
\begin{tabular}{c|c|c|c|c|c|c|c|c|cccc}
\hline
              & \multicolumn{4}{c|}{\textbf{Citeulike}}                                       & \multicolumn{4}{c|}{\textbf{Pinterest}}                                       & \multicolumn{4}{c}{\textbf{Movielens}}                                                                                               \\ \hline
              & \multicolumn{2}{c|}{\textbf{ndcg@k}} & \multicolumn{2}{c|}{\textbf{recall@k}} & \multicolumn{2}{c|}{\textbf{ndcg@k}} & \multicolumn{2}{c|}{\textbf{recall@k}} & \multicolumn{2}{c|}{\textbf{ndcg@k}}                                        & \multicolumn{2}{c}{\textbf{recall@k}}                 \\ \hline
              & \textbf{k=10}     & \textbf{k=20}    & \textbf{k=10}      & \textbf{k=20}     & \textbf{k=10}     & \textbf{k=20}    & \textbf{k=10}      & \textbf{k=20}     & \multicolumn{1}{c|}{\textbf{k=10}}   & \multicolumn{1}{c|}{\textbf{k=20}}   & \multicolumn{1}{c|}{\textbf{k=10}}   & \textbf{k=20}   \\ \hline
Neurec        & 0.029             & 0.0331           & 0.0371             & 0.0468            & 0.0441            & 0.0592           & 0.0537             & 0.088             & \multicolumn{1}{c|}{0.0939}          & \multicolumn{1}{c|}{0.0968}          & \multicolumn{1}{c|}{0.0971}          & 0.1088          \\ \hline
BPR           & 0.0305            & 0.0362           & 0.0368             & 0.0512            & 0.0351            & 0.0472           & 0.0472             & 0.0694            & \multicolumn{1}{c|}{0.0888}          & \multicolumn{1}{c|}{0.0927}          & \multicolumn{1}{c|}{0.093}           & 0.1075          \\ \hline
NGCF          & 0.0398            & 0.0496           & 0.0508             & \underline{ 0.0752}      & 0.0464            & 0.0629           & 0.0563             & 0.0939            & \multicolumn{1}{c|}{0.0899}          & \multicolumn{1}{c|}{0.095}           & \multicolumn{1}{c|}{0.0918}          & 0.1071          \\ \hline
LR-GCCF       & \underline{0.0442}      & \underline{0.0528}     & \underline{ 0.0531}       & 0.0749            & 0.0485            & 0.0635           & 0.0574             & 0.0916            & \multicolumn{1}{c|}{0.0936}          & \multicolumn{1}{c|}{0.0975}          & \multicolumn{1}{c|}{0.0965}          & 0.109           \\ \hline
GCMC          & 0.0384            & 0.0484           & 0.0492             & 0.0737            & \underline{ 0.0495}      & \underline{ 0.0661}     & \underline{ 0.0595}       & \underline{0.0971}      & \multicolumn{1}{c|}{\underline{0.096}}     & \multicolumn{1}{c|}{\underline{ 0.0985}}    & \multicolumn{1}{c|}{\underline{0.1001}}    & \underline{0.1103}    \\ \hline
RH-GCCF         & \textbf{0.0541}   & \textbf{0.0657}  & \textbf{0.0648}    & \textbf{0.094}    & \textbf{0.0544}   & \textbf{0.0716}  & \textbf{0.0646}    & \textbf{0.1037}   & \multicolumn{1}{c|}{\textbf{0.0972}} & \multicolumn{1}{c|}{\textbf{0.1007}} & \multicolumn{1}{c|}{\textbf{0.1011}} & \textbf{0.1136} \\ \hline
Improvement\% & +22.40            & +24.43           & +22.03             & +25.00            & +9.90             & +8.32            & +8.57              & +6.80             & \multicolumn{1}{c|}{+1.25}           & \multicolumn{1}{c|}{+2.24}           & \multicolumn{1}{c|}{+1.00}           & +2.99           \\ \hline
\end{tabular}
\end{table*}
\subsection{Comparison (RQ3)}
\subsubsection{Overall Comparison.}
The performance of baselines and our proposed model are summarized in Table 3. We have the following observations:
\begin{itemize}[leftmargin=10pt]
\item LR-GCCF and NGCF achieves the best performance among baselines on Citeulike; GCMC is the best baseline on Pinterest and Movielens. The consistent improvements over competing baselines across all datasets demonstrate the effectiveness of our model. For instance, the improvement over the best baseline on Citeulike, Pinterest, Movielens is 24.43\%, 8.32\%, 2.24\%, respectively, in terms of ndcg@20. 
\item GCN-based methods perform better on sparse datasets, while MF-based methods (including the deep learning based method) tend to achieve better performance on dense datasets. We speculate that on dense datasets there are enough interactions to describe user preference, where neighborhood messages are redundant and instead introduce useless information. However, MF-based methods can not solve the lack of interactions on sparse datasets. GCN-based methods tackle this by complementing original interactions with neighborhood messages.
\item Among GCN-based methods, GCMC which considers first-order neighborhood performs better on Pinterest, while NGCF and LR-GCCF which consider higher-order neighborhood show superior performance on Citeulike which is sparser than Pinterest. This shows that higher-order neighboor messages do not always leads to better results, for the reason that it's difficult to distinguish the useful messages from the huge number of redundant messages. On the other hand, higher-order neighbor messages enable the model to better comprehend user taste on sparse datasets.
\item Our model achieves better performance on sparser datasets, which is consistent with the above analysis of GCN-based methods. For instance, the improvement of our model over the best baseline is 2.24\% on the relatively denser dataset Movielens, in terms of ndcg@20; while this value is 24.43\% on Citeulike which is the sparsest among the three datasets.   
\end{itemize} 

\subsubsection{Comparison w.r.t Layer-Wise Noise}
To verify if randomly dropping out edges at each layer indeed contributes to performance, we compare our model with the model without randomly dropping out edges, which is illustrated in Figure \ref{layer_comparison}. We have the following findings:
\begin{itemize}[leftmargin=10pt]
\item Our model almost outperforms the 'no-drop' model which puts the whole graph into training across the board, which demonstrates two things. Firstly, the improvements over the model which considers higher-order neighborhood verifies that our model is able to well alleviate over-smoothing and over-fitting. Secondly, our model also performs better than the 'no-drop' when only considering first-order neighborhood, which shows the effectiveness of the sub-graph training strategy.
\item There is a drop on performance when we keep increasing the layers on our model. The reason is twofold. As we mentioned previously, the layer-wise noise is transmitted forward along with neighboor messages, thus the higher-order neighboorhood suffers more from the noise than the lower-order neighborhood; there must be a point after which the higher-order neighbor messages are full of so much noise that instead reduce the performance. What's more, when the neighborhood start to converge which almost contains all node in the graph, keeping increasing neighbor messages would not introduce new information, which only causes over-fitting and increases the training difficulty.     
\end{itemize}

\begin{figure} \centering 
\subfigure[Performance of ndcg@10 on Citeulike with Gaussian noise.  ] {  
\includegraphics[width=0.4\columnwidth]{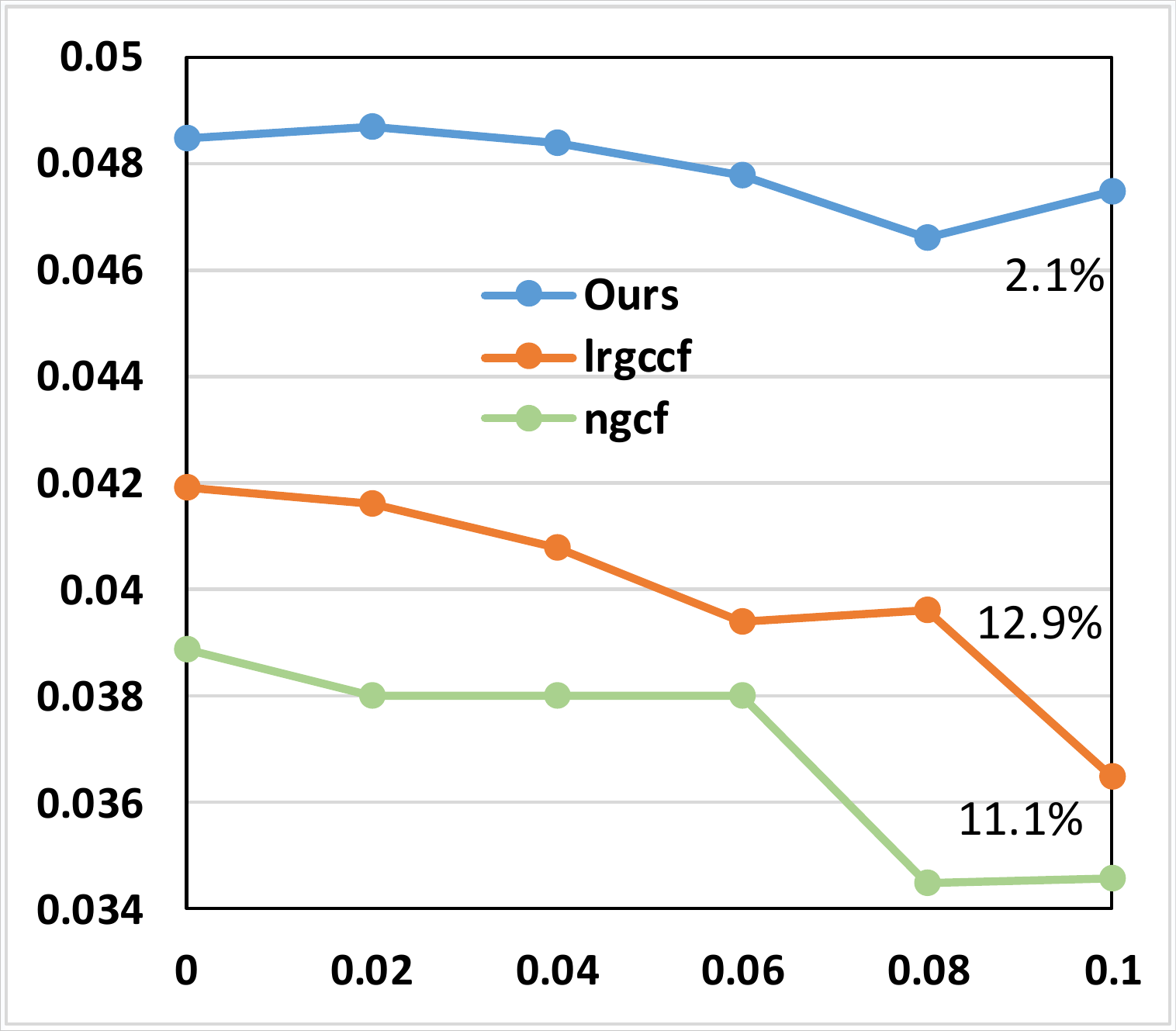} 
}  
\hspace{0.2cm}
\subfigure[Performance of ndcg@10 on Citeulike under data sparseness. ] {  
\includegraphics[width=0.4\columnwidth]{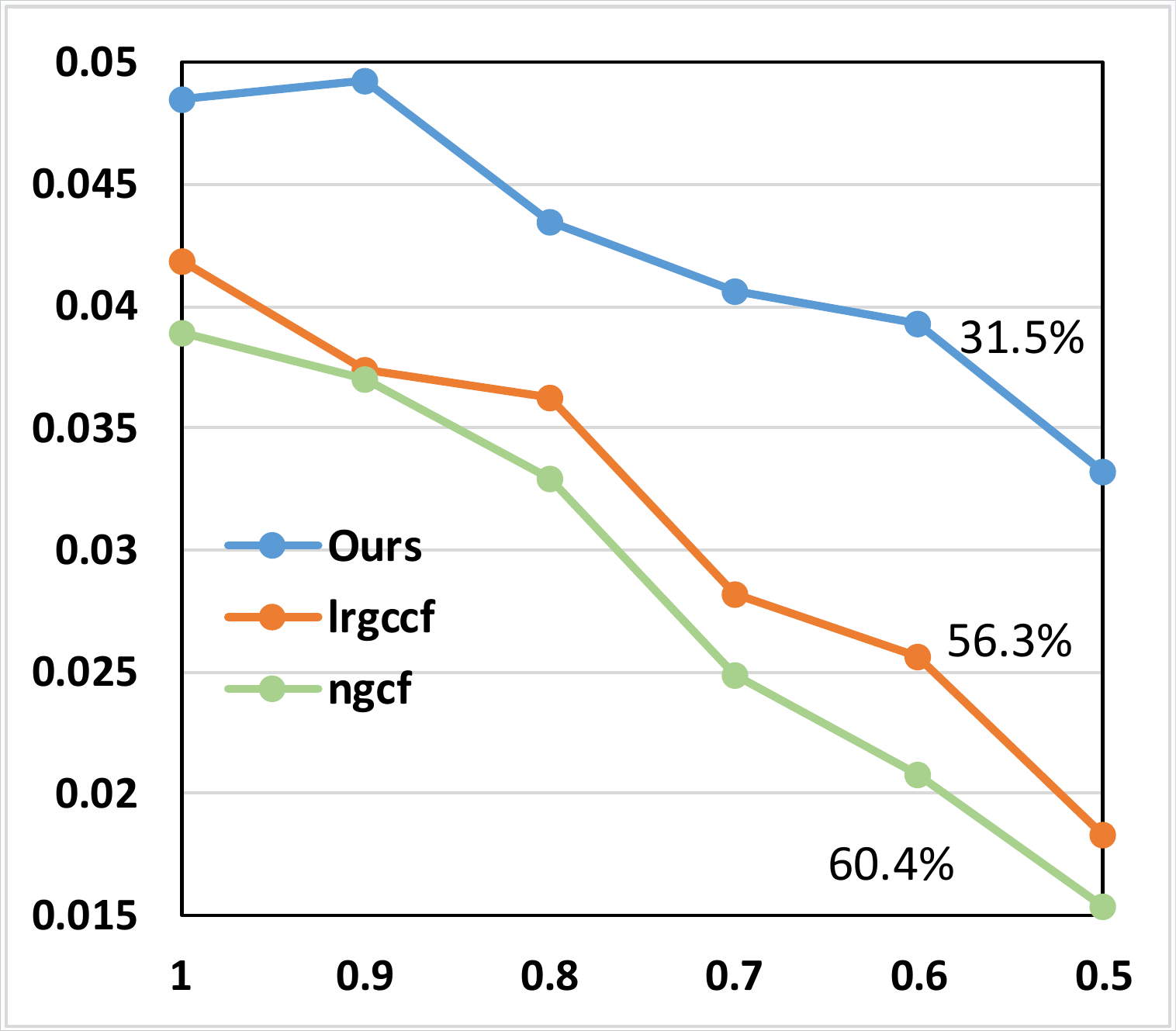} 
}
\hspace{0.2cm}
\subfigure[Performance of recall@10 on Pinterest with Gaussian noise. ] {  
\includegraphics[width=0.4\columnwidth]{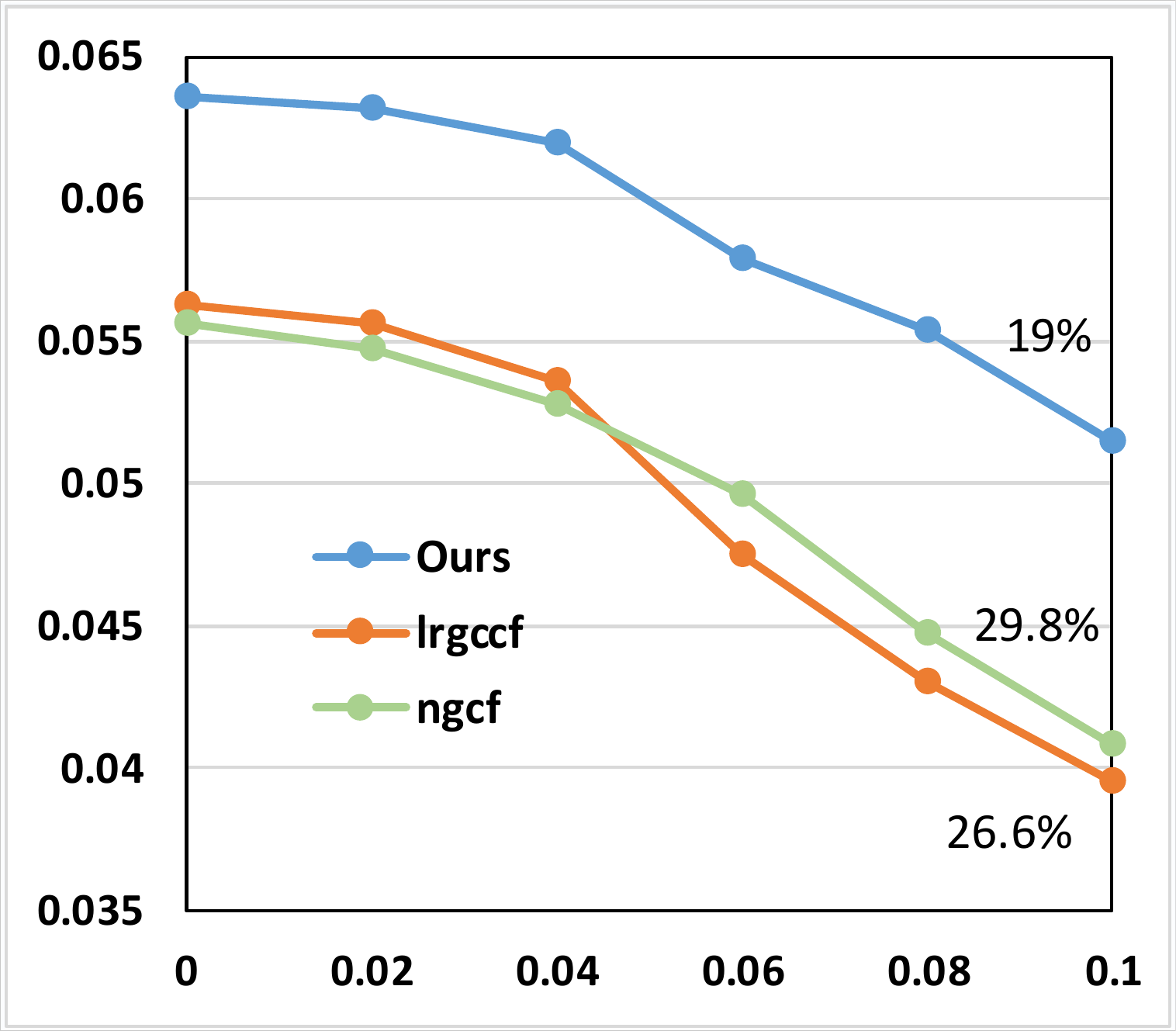} 
}
\hspace{0.2cm}
\subfigure[Performance of recall@10 on Pinterest under data sparseness. ] {  
\includegraphics[width=0.4\columnwidth]{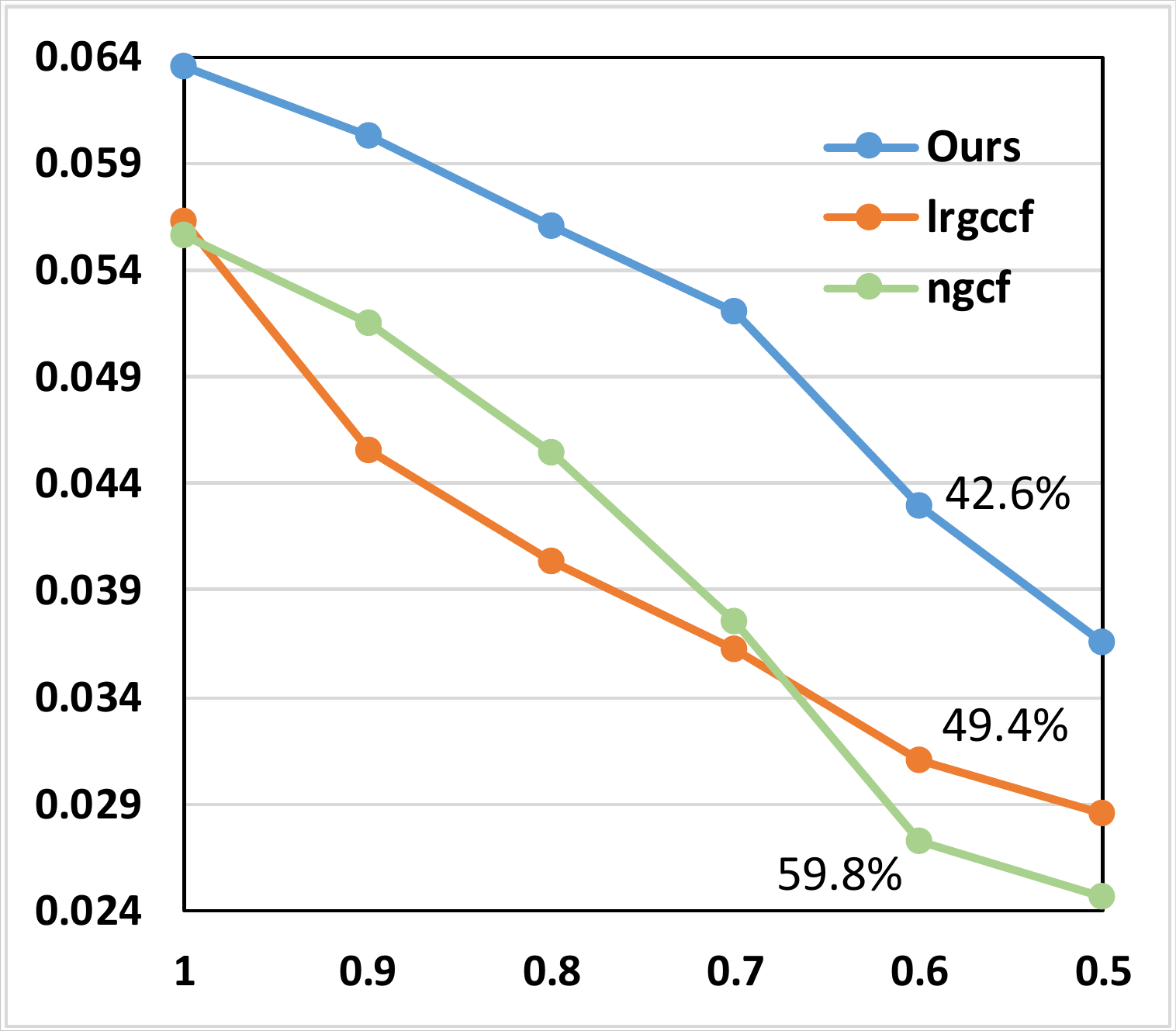} 
}   
\caption{ Performance under two kinds of adversarial attacks ($K=2$). }  
\label{attacks}
\end{figure}

\begin{figure}
\begin{center}
\includegraphics[width=0.22\textwidth]{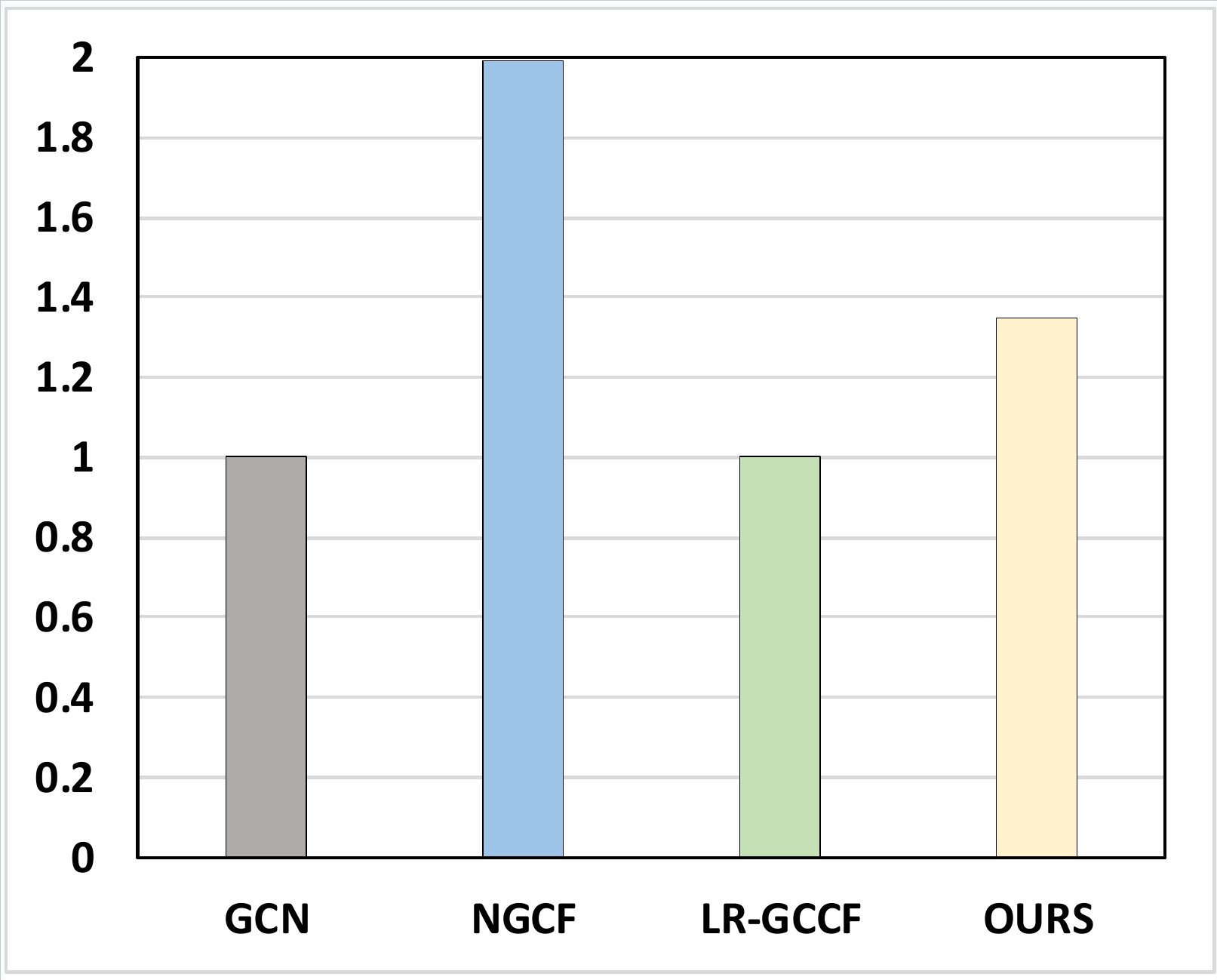}
\caption{Comparison of training time per epoch.}
\label{time}
\end{center}
\end{figure}

\subsubsection{Comparison w.r.t Adversarial attacks}
We conduct experiments to test the robustness of GCN-based methods with two common adversarial attacks:
\begin{itemize}[leftmargin=10pt]
\item For node representations per layer, we add zero-mean Gaussian noise $\mathcal{N}(0,\sigma^2)$. We change the standard deviation to adjust the intensity of the noise.
\item Each edge connection is randomly dropped out with $1-p$, we change $p$ to adjust the intensity of the noise.
\end{itemize}
Figure \ref{attacks} shows the performance under adversarial attacks and the degradations compared to the model under the most intense attacks. NGCF is more robust than LR-GCCF under random noise; while LR-GCCF performs better under data sparseness, which may because of the lower model complexity. Our model achieve consistent improvements as well as lower degradations over the other two methods, which demonstrates that our model is able to offer robust recommendations under different adversarial attacks.      
\subsubsection{Comparison w.r.t Training Time.}
Figure \ref{time} reports the training time of several GCN-based methods. For simplicity, we set the training time of GCN as the benchmark, and for the sake of fairness all models are set to three layers. We can see that LR-GGCCF has the lowest model complexity, on account of the linear embedding
propagation; while NGCF is the most time-consuming model, where introducing additional model parameters increase the model complexity. The additional complexity of our model comes from the binary matrix (vector), where the additional runtime is about 0.34 times of the training time of GCN, which is acceptable considering the improvement over other GCN-based methods.

\section{Conclusion}
In this paper, we proposed a robust hierarchical graph convolution network for collaborative filtering (RH-GCCF), which aims at improving GCN for robust recommendations. We first proposed a solution for over-smoothing and over-fitting by randomly dropping out node messages at each layer, which shares similarities with dropout \cite{srivastava2014dropout}. Then we built a hierarchical model by separately aggregating node messages from different order-neighborhood, which avoids mixing them indistinguishably. We conducted extensive experiments on three real-world datasets to evaluate our proposed model, regarding the performance, complexity and robustness. The experimental results verifies effectiveness and robustness of our proposed model. In future, we are committed to representing higher-order neighborhood in a more reasonable and effective way for better comprehension of user preference and robust recommendations. 
                                

\bibliographystyle{ACM-Reference-Format}
\bibliography{sample-base}


\end{document}